\documentclass[a4paper,fleqn, review]{cas-sc}

\usepackage[numbers,sort&compress]{natbib}
\usepackage{placeins}
\usepackage{bbm}
\usepackage{graphicx}
\usepackage{longtable}
\usepackage{pdflscape}
\usepackage{longtable}

\newcommand{\red}[1]{\textcolor{red}{#1}}

\begin{document}
\let\WriteBookmarks\relax
\def\floatpagepagefraction{1}
\def\textpagefraction{.001}

\shorttitle{The rise of science low-carbon energy technologies}
\shortauthors{H\"otte, Pichler, Lafond 2020}

\title[mode = title]{The rise of science in low-carbon energy technologies}

\author[1,2]{Kerstin H\"otte}[orcid=0000-0002-8633-4225]
\cormark[1]
\author[3,4,5]{Anton Pichler}[orcid=0000-0002-7522-1532]
\author[3,4]{Fran\c{c}ois Lafond}[orcid=0000-0002-8333-561X]


\address[1]{Faculty of Business
Administration and Economics, University of Bielefeld, D-33615, Germany}
\address[2]{CES-Centre d'\'{E}conomie de la Sorbonne, Universit\'{e} Paris 1 Sorbonne Panth\'{e}on, Paris Cedex 13, France}
\address[3]{Institute for New Economic Thinking at the Oxford Martin School, University of Oxford, OX26ED, UK}
\address[4]{Mathematical Institute, University of Oxford, OX13LP, UK}
\address[5]{Complexity Science Hub Vienna, 
A-1080, Austria}

\cortext[cor1]{Corresponding author: kerstin.hoette@uni-bielefeld.de}

\begin{abstract}
\noindent 
Successfully combating climate change will require substantial technological improvements in Low-Carbon Energy Technologies (LCETs), but designing efficient allocation of R\&D budgets requires a better understanding of how LCETs rely on scientific knowledge.
Using data covering almost all US patents and scientific articles that are cited by them over the past two centuries, we describe the evolution of knowledge bases of ten key LCETs and show how technological interdependencies have changed over time. The composition of low-carbon energy innovations shifted over time, from Hydro and Wind energy in the 19th and early 20th century, to Nuclear fission after World War II, and more recently to Solar PV and back to Wind. In recent years, Solar PV, Nuclear fusion and Biofuels (including energy from waste) have 35-65\% of their citations directed toward scientific papers, while this ratio is less than 10\% for Wind, Solar thermal, Hydro, Geothermal, and Nuclear fission. Over time, the share of patents citing science and the share of citations that are to scientific papers has been increasing for all technology types. The analysis of the scientific knowledge base of each LCET reveals three fairly separate clusters, with nuclear energy technologies, Biofuels and Waste, and all the other LCETs. 
Our detailed description of knowledge requirements for each LCET helps to design of targeted innovation policies.
\end{abstract}

\begin{highlights}
\item Patents and citations to science are used to analyze the past 200 years of low-carbon energy innovation. 
\item Low-carbon energy inventions became increasingly science-based. 
\item Non-fossil fuels, Nuclear and Solar energy are most science-intensive, Hydro power is the least. 
\item Nuclear, Fuels, and Renewables form three fairly separate knowledge clusters.
\end{highlights}

\begin{keywords}
Low-carbon energy \sep 
Patents \sep
Citation networks \sep 
Non-patent literature \sep
Science-technology relationship \sep
History of technology \sep
Reliance on science
\end{keywords}

\maketitle
\textbf{Word count:} 7394 
\section*{Abbreviations}
\begin{description}
\setlength{\itemsep}{-5pt} \small
\item [CPC] Cooperative Patent Classification
\item [CS] Confidence Score
\item [ECLA] European Classification System
\item [EPO] European Patent Office
\item [IPC] International Patent Classification
\item [LCET] Low-Carbon Energy Technology
\item [MAG] Microsoft Academic Graph
\item [R\&D] Research and Development
\item [RoS] Reliance on Science
\item [USPTO] United States Patent and Trademark Office
\item [WoS] Web of Science
\end{description}

\section{Introduction}

To meet the Paris Agreement and to reduce climate risks, the transition to Low-Carbon Energy Technologies (LCETs) needs to be accelerated \citep{galiana2009let, rogelj2016paris, ipcc2018special}. R\&D can stimulate the invention and diffusion of LCETs \citep{popp2016economic, pless2020bringing, anadon2017integrating, farmer2019sensitive}, but it is not clear whether the main driver of technological progress is fundamental science, applied science or applied technological development \citep{corsatea2014technological, kline2010overview,meyer2000does, lacerda2020effectiveness}. Since the drivers of progress differ across technology types \cite{mansfield1995academic, verbeek2002linking}, a systematic and technology-specific understanding of these drivers is essential to design targeted and efficient innovation policies. 
To this aim, we look at the patterns of citations of LCET patents over two centuries, and describe quantitatively and qualitatively how the relative importance of different knowledge resources differs across technology types, and how they evolved over time.

Over the past decade, an increasing number of studies using patent data has documented that green inventions are more novel, more radical and have a more pervasive effect on subsequent innovation compared to non-green alternatives \citep{noailly2013technologies,barbieri2020knowledge,dechezlepretre2011invention,noailly2017knowledge}. This suggests that the shift of innovative efforts from fossil to low-carbon technologies can be a trigger of positive innovation dynamics and growth.
But which LCET is the most promising? And should innovation policy target fundamental science or applied technological development?
While it is well acknowledged that public research institutions, often oriented toward more fundamental science, play a central role in the invention of LCETs \citep{quatraro2019academic, triguero2013drivers, orsatti2019public, popp2019environmental, lacerda2020effectiveness}, it remains poorly understood how this role differs across technologies and over time.

Following Brian Arthur's definition \citep{arthur1989competing}, technologies are ``a collection of phenomena captured and put to use''. Different ``families of [scientific] phenomena" (chemical, electrical, quantum, etc.) lead to distinct families of technologies based on these phenomena. 
Intuitively, it is known that LCETs rely on very different scientific principles or physical phenomena, and on rather distinct fields of science. Nuclear fusion is based on the physics of plasmas, Hydro energy on mechanical forces, Biofuels on chemistry and Solar photovoltaics (PV) on the photoelectric effect. 
In this article, this theoretical conception of technological heterogeneity is systematically brought to data. 

Interactions of science and technology work in both directions, and there are strong interdependencies among different technological and scientific domains \citep{jaffe2019patent,small2014identifying,meyer2000does,verbeek2002linking, tijssen2001global}. This makes it difficult to map the exact relationships between specific technologies and specific technological domains.
A systematic, data-driven analysis could build on the study of patent citations to the ``non-patent literature'', which is often of scientific origin \citep{narin1995linkage}. However, it is only recently that high-quality and comprehensive data became available. Analyses at the aggregate level have shown that most patents are linked to a scientific paper and vice-versa, typically through a path of 2-4 citations \citep{ahmadpoor2017dual}. Two recent studies, using slightly different methods and data, found that patent value is related to the quality of the science it relies on \citep{watzinger2019standing,poege2019science}.

A recent, related study based on data from the European Patent Office (EPO) found that renewable energy technologies rely on more recent, more diverse and more frequently cited scientific literature, compared to fossil-fuel based technology \citep{persoon2020science}, but did not investigate the evolution of these patterns over time. Another study \citep{albino2014understanding} found that the number of citations to scientific literature increased for ``alternative energy production`` and Nuclear power patents, but did not look at single technologies and the qualitative nature of citation links. 

The analysis reported here expands the study of science-technology interactions for LCET development to a historical scale. It is based on a significantly larger data set covering a much higher number of patents and citation links. 
The size of the data set enables a systematic comparison of different LCETs among themselves and over time, and a closer analysis of the content of the scientific literature cited by LCET patents.

\begin{figure}
\centering
\includegraphics[width = .75\textwidth, trim = {2.5cm 5cm 3cm 4cm}, clip]{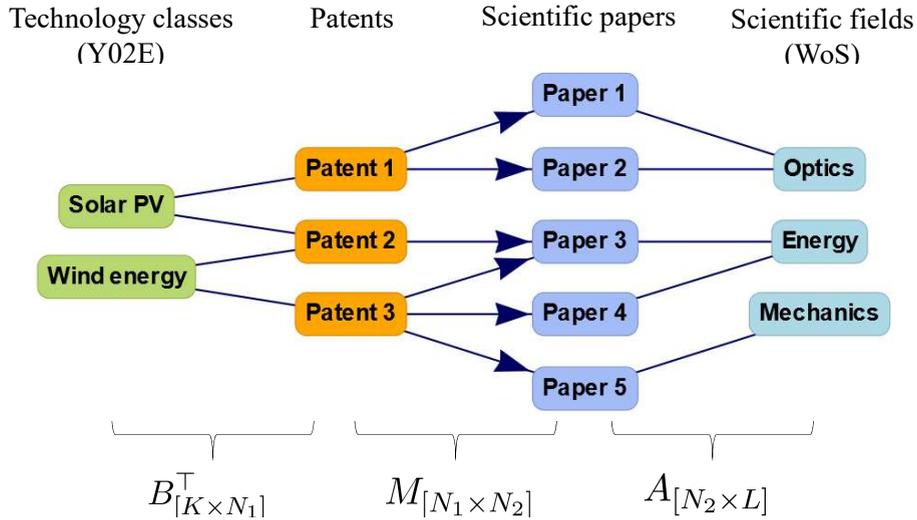}
\caption{
Relationships between technological classes, patents, scientific papers and scientific fields as ``network of networks''. The matrices $B$ and $A$ denote the bipartite patent (row) - technology (column) and paper (row) - field (column) relationships, respectively. Multiple technology classes can be assigned to a patent, but there is only one scientific field (Web of Science) per paper. 
The matrix $M$ couples the technological and scientific layers where a matrix element $M_{ij}=1$ if patent $i$ cites paper $j$ and zero otherwise.
}
	\label{fig:nw_scheme}
\end{figure}

Specifically, this paper studies the evolution of citation links between virtually \emph{all} patents filed at the United States Patent and Trademark Office (USPTO) since 1836\footnote{
The USPTO was created in 1790 but a fire in 1836 destroyed most existing records.
} and published scientific work (Fig. \ref{fig:nw_scheme}). 
The analysis disaggregates ten technology types: Solar PV, Wind energy, Solar thermal (``Thermal'' for short), Ocean power, Hydro energy, Geothermal, Biofuels, Fuels from waste, Nuclear fission, and Nuclear fusion. The data set (Reliance on Science (RoS)), which was only recently assembled, links Microsoft Academic Graph (MAG) data with USPTO patents \citep{marx2019reliance, sinha2015overview, wang2019review}. 
US patents are well suited for comprehensive historical studies, due to the extensive time coverage and high quality of the historical records. Moreover, the US plays a leading role in green patenting \citep{albino2014understanding}, and valuable patents tend to be filed in more jurisdictions \citep{harhoff2003citations} so that important patents of non-US origin are likely to be filed in the US too.

The data covers more than 179 million academic papers classified into 251 scientific fields, 10 million patents classified into more than 240,000 highly detailed technology categories, and 34 million citations from more than 2 million unique patents to 4 million papers (Table \ref{tab:overview_data}).
This paper focuses on a subset of 65,000 patents assigned to 10 different technologies accounting for all low-carbon electricity generation\footnote{%
These technologies, especially Biofuels and Geothermal, may also be used for producing other forms of energy such as heat.
} (Table \ref{tab:electricity_gen}) and describes their interactions with scientific knowledge across time.

\begin{table}[!t]
\centering
\begin{tabular}{p{4cm}p{2cm}|p{4cm}p{2cm}}
\hline
\textbf{Full data set}&&\textbf{LCET subset}&  \\ 
\hline
\# papers in MAG           & 179,083,029&&\\
\# patents                 & 10,160,667 & \# patents           &65,305\\
\# citing patents          &  2,058,233 & \# citing patents    &22,017\\ 
\# cited papers            &  4,404,088 & \# cited papers      &103,645\\
\# citation links          & 34,959,193 & \# citation links    &396,504\\ 
\hline
 \multicolumn{4}{l}{\underline{\textbf{Meta-information:}}}                       \\ 
\multicolumn{4}{p{12cm}}{\textbf{Papers:} Publication year, 251 Web-of-Science (WoS) categories, Journal/ conference proceedings name, DOI, Paper title}\\
\multicolumn{4}{p{12cm}}{\textbf{Patents} Grant year, $>$250,000 hierarchical CPC classes, 10 LCET types}\\
\multicolumn{4}{p{12cm}}{\textbf{Citation links:} Reference type, citation type, reliability score}\\
\hline
\end{tabular}
\caption{Overview of data. The LCET subset contains ten different technologies: Solar PV, Wind energy, Ocean power, Hydro energy, Geothermal, Solar thermal, Biofuels, Fuels from waste, Nuclear fission, Nuclear fusion.}
\label{tab:overview_data}
\end{table}

\begin{table}[]
\begin{tabular}{|l|l|l|l|}
\hline
\textbf{Energy technology} & \textbf{TWh} & \textbf{Share total (\%)} & \textbf{Share LCET (\%)} \\
\hline
Solar PV          & 443.55     & 1.72             & 4.88            \\
Wind energy      & 1,127.32   & 4.38             & 12.39           \\
Solar thermal     & 10.85      & 0.04             & 0.12            \\
Ocean power       & 1.04       & 0.00             & 0.01            \\
Hydro energy      & 4,197.30   & 16.32            & 46.14           \\
Geothermal        & 85.35      & 0.33             & 0.94            \\
Biofuels          & 481.53     & 1.87             & 5.29            \\
Fuels from waste  & 114.04     & 0.44             & 1.25            \\
Nuclear fission   & 2,636.03   & 10.25            & 28.98           \\
Nuclear fusion    & 0.00       & 0.00             & 0.00            \\
\hline
\textbf{Total LCET}        & 9,097.01   & 35.37            & 100             \\
\hline
Coal              & 9,863.34   & 38.35            &                 \\
Gas               & 5,882.83   & 22.87            &                 \\
Oil               & 841.88     & 3.27             &                 \\
Other sources     & 36.02      & 0.14             &                 \\
\hline 
\textbf{Total All}             & 25,721.08  & 100              & \\
\hline
\end{tabular}
\caption{Global electricity generation per technology type in 2017. Source: \url{https://www.iea.org/data-and-statistics}{}.} \label{tab:electricity_gen}
\end{table}

As in \citep{persoon2020science}, we find that some LCETs are vastly more science-intensive than others.
The scientific dependence of all LCETs has increased substantially over time reflecting the evolution from individual inventors in the 19th century to industrial systems of innovation. In terms of composition, LCET patenting shifted from Hydro and Wind, to Nuclear, and to Solar PV and Wind recently. Looking at the reliance on science, four key findings emerge: 1) All LCETs increasingly rely on science. 2) Biofuels and Fuels from Waste rely most heavily on science which is due to their close connection to biochemical research. Solar PV and Nuclear fusion also rely heavily on science, in sharp contrast to other technologies, particularly Hydro. Science-intensive technologies tend to rely more strongly on basic rather than applied science. 
3) Looking at the qualitative nature of the knowledge base, the two fuel-based, the six non-fuel-based Renewables and the two Nuclear energy technologies form separate clusters that rely on a strikingly different kind of science. 
4) The scientific bases of Renewables have become more similar over time, suggesting synergies in technological development and research.
This is also true for fuel-based technologies, but not for the two nuclear power technologies.
The results suggest that the degree to which public funds should be directed to basic or applied science rather than applied technological development is, first, dependent on the technology type that is targeted, and second, likely to be higher now than it was in the past.

There are several limitations of our study, such as potential biases due to the fact that different fields and different periods have different norms and practices for patenting, publishing, and making citations. Our analysis is also limited to the development stage of technologies and does not account for drivers and barriers at later stages of LCET deployment, diffusion and up-scaling which are also relevant for policy. Nevertheless, our analysis draws a detailed but also systematic picture of the knowledge base of low-carbon energy, showing which types of knowledge resources are required to develop different types of LCET. 

Technology-specific support policies may focus on the provision of these resources, distinguishing between the support for applied development, applied and basic sciences, and different qualitative fields of research such as bio-chemistry, mechanical engineering  or plasma physics. Moreover, the network perspective helps to compare alternative technological pathways. The convergence observed in Renewables indicates that these technologies increasingly benefit from mutual spillovers of R\&D activities. This can be a significant leverage mechanism that amplifies R\&D investment in these domains. 

The paper is organized as follows. Section \ref{methods} describes our methods, Section \ref{results} presents the results, and Section \ref{conclusion} offers a discussion of limitations and policy conclusions.

\section{Methods and data}
\label{methods}

\subsection{Data sources}

The major input to this analysis is the publicly available data set Reliance on Science (RoS) data set\footnote{This study uses Version 23 [Feb 24, 2020]: \url{https://doi.org/10.5281/zenodo.3685972}} that links scientific papers of MAG and a subset of the available meta-information on papers \cite[]{marx2019reliance, sinha2015overview, wang2019review}. 
This data is complemented by a second data set that provides detailed information on USPTO patents \citep{pichler2020technological}. 

An overview of the data is provided in Table \ref{tab:overview_data} and additional statistics are summarized in Appendix \ref{si:stats_on_subset}. 
To ensure reproducibility, all data is provided in an online data repository \citep{hotte2020data}.\footnote{\url{https://doi.org/10.4119/unibi/2941555}}

\subsubsection{Reliance on Science}
\label{sec:methods_data_ros}
RoS covers more than 2 million unique patents with 34 million citation links to 4 million unique scientific articles in a time horizon from 1800 (1836) for the earliest paper (patent) until 2019. 
The data is complemented by data on the publication year of papers and grant year of patents. This allows us to study the evolution of the number of patents and papers over time. 

The developers of RoS used a statistical procedure to match non-patent references made in patent documents to unique papers in MAG and provide confidence scores ($CS\in \{1,10\}$) for the reliability of matching links. Links with a $CS>3$ are used, which is associated with a precision rate of $98.76\%$ and a recall of $93.01\%$ \cite[cf.][]{marx2019reliance}. Approximately 75\% of the patents in our data set have the highest confidence score (Appendix, Table \ref{tab:gen_stats}) and results shown here are robust with respect to confidence scores (Appendix \ref{sec:SI_robust_CS}). General statistics about the citation links in our data set are summarized in Appendix \ref{si:stats_on_subset}.

The data includes unique identifiers for scientific papers that can be associated with meta-information about the publication (e.g. citations to and from other papers, year of publication, scientific discipline, journal or conference publication, impact factors, DOIs, authors and their affiliations). We use only direct paper citations, publication years and scientific fields classified into 251 Web-of-Science (WoS) categories. 
	
The publication year can be interpreted as an approximate time stamp for the first discovery of the scientific knowledge that is encoded in the research article. Some patents cite early versions of papers that were published more formally 1-2 years later. Multiple versions of papers are improperly distinguished in the data and typically only the published version is reported. This occurs, for example, when an early version of a paper is available as pre-print and a patent is citing this pre-print prior to the publication of the paper in a journal. This can lead to negative citation lags because papers in MAG usually refer to the journal publication. This inaccuracy of citation lags is negligible given the time scale of 200 years and because it does not alter the other characteristics of the citation link. 

WoS categories are commonly assigned to papers at the level of journals. Normally, the assignment is done by Clarivate Analytics and follows a heuristic procedure mixed of algorithmic procedures based on existing categorizations and expert opinion \citep{wang2016large}. Clarivate Analytics is a company specialized in the supply of data-related services and analytics of intellectual property and science. Since 2016, it maintains the WoS database on scientific citation data. 
In the RoS data, the WoS categories are assigned to papers at the paper level using the more than 200,000 different categorization tags called ``Field of Study''. These tags have been automatically generated by machine learning routines in MAG \citep{sinha2015overview, wang2019review}. \citet{marx2019reliance} have probabilistically matched MAG fields to WoS categories and offer a direct mapping of papers to WoS categories. 

The titles of papers and DOIs allow to trace a citation made in a patent document back to the data at the micro-level. Hence, individual citation links can be inspected manually looking at the paper and the patent document. Some implausible citation links were checked manually, e.g. if a patent maps to an unexpected field in WoS (e.g. if Solar PV cites "Poetry") or if a patent cites a paper with a more recent publication date than the patent grant year. Such false-positives have been removed manually.\footnote{
The code used for the data compilation in the data publication is provided only to ensure transparency and reproducibility \cite{hotte2020data}.
}

\subsubsection{Patent data}
\label{sec:methods_data_pat}

We use a database recently put together by \citet{pichler2020technological}. It merges various sources including the USPTO bulk downloads and PatentsView, Google Patents Public Data (provided by IFI CLAIMS Patent Services), and citations data from \citet{kogan2017technological} and \citet{lafond2019long}. It contains all utility patents granted by the US patent office (USPTO) from 1836 to early 2019, except for rare exceptions, as well as most available citation data.

The year of the patent refers to the publishing (granted) date, since application dates for older patents are often not available.
To identify different technological categories, we used the Cooperative Patent Classification (CPC). The CPC system was developed by the EPO together with the USPTO to harmonize patent classifications and to replace the former European Classification System (ECLA) and US patent classification (USPC) systems. The CPC is similar to the International Patent Classification (IPC) system, but is more detailed and appears more thoroughly applied, particularly to old patents. 

It is important to note that the classification system changes through time to reflect the evolution of technology. Patent offices reclassify old patents into the current classification system to ensure that the classification of patents is always up-to-date. This makes it possible to search for prior art and to evaluate the novelty of patent applications. The reclassification of old patents is done automatically by the USPTO and it cannot be guaranteed that the data is free from misclassifications. To address this problem, we checked manually the oldest patents by looking at the patent title. In case of inconsistent or implausible titles, we checked the patent documents themselves. We did this for all LCET patents granted prior to 1865 and, additionally, for the 10-20 oldest patents of each technology class. 

Two observations emerged from this manual validation: (1) The majority of old patents is classified correctly and refers to inventions that have a direct and obvious relation to the energy technology (e.g. patents for water mills are classified as Hydro energy patents). (2) In some cases and in particular for more complex technologies that combine different components, this relation is less clear. For example, patents for the improvement in reflection techniques of solar light used for photography are classified as patents for Solar thermal energy. Some patents for milk churns are classified as Hydro energy patents which is likely due to their rotating paddle used to churn milk. These relationships to Solar thermal and Hydro energy seem plausible, and technological boundaries are evolving, so that deciding what constitutes a misclassification error would be arbitrary. 

Because the number of patents with an unclear classification is small, and to avoid introducing subjectivity, these patents were kept in the data set\footnote{
An exception is a patent from the 1830s granted for a method to make forks and knives and classified as Nuclear patent, which we removed from the data set used to produce Fig. \ref{fig:timeseries_shares}.
}.
The continuous reclassification of patents also means that future versions of the patent data set will classify differently some of the patents that are analyzed here \citep{lafond2019long}. 

The CPC classifications include "Y02E" tags, indicating climate change mitigation technologies ``related to energy generation, transmission and distribution''. Y-tags do not form separate technological classes, but are additional tags attached to patents by examiners. Since climate change mitigation technologies tend to develop extremely fast and are distributed over a wide range of technological fields, patent offices introduced a separate tagging scheme \citep{veefkind2012new}. 
This analysis focuses on a set of ten LCETs which can be identified by Y02E tags: Solar PV, Wind, Solar thermal, Ocean power, Hydroelectric, Geothermal, Biofuels, Fuels from Waste,  Nuclear fission and Nuclear fusion.

\subsection{Subsetting procedure and data-cleaning} 
Beginning with the comprehensive data set of $10$ million patents and 179 million academic papers, two data subsets of LCET patents and LCET patents that cite to science were sequentially extracted. Along the sequence of subsetting steps, the data was cleaned from incomplete or inaccurate entries and complemented with different types of meta-information. 

	\paragraph{Step 1: Subset green energy technologies:} From the complete list of USPTO patents covering the time horizon from 1836 to 2019, those patents were selected that have 6-digit CPC codes that are associated with Renewable (\texttt{Y02E10}), Non-fossil fuels (\texttt{Y02E50}) and Nuclear energy (\texttt{Y02E30}). 
	\paragraph{Step 2: Disaggregate LCET by technology type:} 
	The aggregate LCET data is split into ten distinct categories identified by 7-digit CPC codes (i.e. \texttt{Y02E10/1} (Geothermal), \texttt{Y02E10/2} (Hydro), \texttt{Y02E10/3} (Ocean), \texttt{Y02E10/4} (Solar thermal), \texttt{Y02E10/5} (Solar PV), \texttt{Y02E10/7} (Wind), \texttt{Y02E50/1} (Biofuels), \texttt{Y02E50/3} (Fuels from Waste), \texttt{Y02E30/1} (Fusion), \texttt{Y02E30/3} and \texttt{Y02E30/4} (Fission). 
	
	Two issues occurred during the data processing: (1) Some patents are classified by coarse CPC codes indicating that these patents are useful for multiple technology types. For example, a patent tagged by \texttt{Y02E10/00} is useful for all types of Renewable energy. These multiple-purpose patents are counted repeatedly, once for each technology type. 
	This splitting procedure is applied to 7 patents tagged with \texttt{Y02E10/00} (all Renewables excluding Biofuels and Waste), 256 patents with \texttt{Y02E10/60} (Solar thermal and PV), 3 patents with \texttt{Y02E50/00} (Biofuels and Waste), 2 patents with \texttt{Y02E30/00} (Fusion and Fission). 
	(2) Patents may have two CPC tags within the same 8-digit class. This situation arises in some cases when applying the allocation rule for multiple-purpose technologies and it arises when a patent has multiple 8-digit classes within one 7-digit class, for example \texttt{Y02E10/32} and \texttt{Y02E10/33}. These double counts are not intended are removed from the data set. 
	
	The final subset comprises 65,305 unique patents and 69,664 patent entries when multiple-purpose patents are counted repeatedly. 
	\paragraph{Step 3: Matching of LCET patents to paper citations:}
	This LCET data set is further subset to obtain a second data set on LCET patents that cite science.
	\citet{marx2019reliance} have matched patents to scientific articles and offer a file that provides a direct mapping between patent numbers and paper identifiers (ID). One patent maps to multiple papers if more than one paper is cited. Another data set is compiled that maps the ten types of LCETs to scientific articles. Patents that do not cite scientific papers are removed from this second data subset.
	\paragraph{Step 4: Complementing with meta-information:} This second data subset of LCET-science citations is complemented by meta-information about the scientific papers that are cited. The meta-information includes the year of publication, the WoS category, paper and patent title. Citations to papers that cannot be assigned to a scientific field are removed from the data set. Whenever available, the DOI, journal or conference name are added. The data also comprises information about the type of citation, i.e. whether the citation was added by the patent applicant or examiner, whether the citation was made on the front page or in the text body of the patent, and a confidence score about the reliability of the citation link. Note that the data by \citet{marx2019reliance} was generated by a probabilistic matching procedure. During the research for this article, some false-positive patent-to-paper links were discovered and removed manually. 
	
	The final subset contains 396,504 citation links from 22,017 unique patent assigned to 10 different LCETs to 103,645 unique scientific papers belonging to 235 different scientific fields. 

\subsection{Reliance on science and similarity networks} 
\label{sec:methods_networks}
To quantify similarity between LCETs, the network interpretation of relationships between technological classes, patents, scientific papers and scientific fields is used (Fig. \ref{fig:nw_scheme}).

Consider $K$ technological classes which are assigned to $N_1$ patents which cite scientific articles.
Then the $N_1 \times K$ bipartite technology-patent network can be defined as $B_{ik} =1$ if a technology class $k$ is assigned to patent $i$. 
Similarly, the $N_2 \times L$ matrix with elements $A_{jl}=1$ if the scientific field $l$ is assigned to paper $j$, and zero otherwise, can be defined where $N_2$ is the number of cited papers and $L$ the number of separate scientific fields.
Patents are frequently classified with multiple CPC codes. In contrast, only one WoS field is attached to a paper.\footnote{
WoS fields are originally assigned to journals in the original WoS indices \citep{wang2016large}. However, in the RoS data, MAG field IDs have been used to classify individual papers by field of research which were then mapped to WoS fields. This allows us to retrieve scientific fields for papers that are not listed in WoS.
}
Thus, every row in the paper-scientific field matrix $A$ contains only one non-zero element, whereas there can be multiple 1's per row in the patent-technological class matrix $B$. 
The scientific and technological spaces are coupled via citations going from patents to papers, which is naturally represented by a binary $N_1 \times N_2$ citation matrix $M$. Note that a patent may cite multiple papers.

The $k^{\text{th}}$ row of the matrix $B^\top M$ gives the number of citations from technology class $k$ to any paper $j$. In the same manner, the $k^{\text{th}}$ row of the projection $\tilde{O} = B^\top M A$ gives the number of citations from technology class $k$ to any scientific field $l$\footnote{
As suggested by \cite{pichler2020technological}, matrix $B$ could be row-normalized for the matrix projections to downsize contributions of backward citations made by patents which are assigned to many different classes. The results are very similar for both cases and the binary matrix $B$ is used.
}. 
The share of all citations from technology $k$ going to scientific field $l$ is then given by the row-stochastic matrix $O$ with elements $O_{kl} = \tilde{O}_{kl} / \sum_{l=1}^L \tilde{O}_{kl}$. 
Fig. \ref{fig:green_reliance_on_wos} is constructed using the four largest values for every row in $O$. A more detailed decomposition of LCETs' scientific knowledge is shown in Appendix, Fig. \ref{fig:SI_science_reliance_change}.

Similarities between scientific knowledge bases of technologies are quantified by computing cosine similarities between LCETs' reliance on scientific fields. Since the shares of scientific fields in citations are embodied in the matrix $O$, the pairwise cosine similarities of two technologies' scientific reliance can be computed as $ O_{k_1*} O_{k_2*}/ (||O_{k_1*}|| \, ||O_{k_2*}||)$, where $O_{k*}$ denotes the $k^{\text{th}}$ row of matrix $O$.

By using citations to patents, similarities between LCETs based on their technological instead of their scientific knowledge bases can be quantified as well. Let us denote the patent-patent citation network by the $N_1 \times N_1$ matrix $H$ where the element $H_{ij} = 1$ if patent $i$ cites patent $j$ and zero otherwise. In contrast to the bipartite patent-science citation network $M$, the patent citation network $H$ is square, since rows and columns are both populated by patents. Nevertheless, the algebra above for constructing technology networks can be easily extended. 

In a similar way as before, the total number of citations from one technology to other technologies can be obtained by the projection $\tilde{P} = B^\top H B$. The row-normalization $P_{k_1 k_2} = \tilde{P}_{k_1 k_2}/ \sum_{l=1}^L \tilde{P}_{k_1 l}$ then gives the distribution of a technology's citation over all other technological classes. Taking pairwise cosine similarities of all rows of matrix $P$ yields a measure of similarity between LCETs' technological knowledge base (instead of to the scientific knowledge base considered above).

To check robustness, more fine-grained bibliographic networks are computed as alternative measures of knowledge base similarities. Here, two LCETs are similar if they cite the same papers (instead of scientific fields) or if they cite the same patents (instead of technological classes).
Details on the derivation of bibliographic coupling networks, as well as results, are presented in Appendix \ref{sec:SI_bibcoup}.

\section{Results}
\label{results}

\subsection{Innovation dynamics of low-carbon energy technologies}
The focus of this analysis is on LCETs, including Nuclear fission and fusion and non-fossil fuels, which, as will be shown, differ not only by their scientific base but also by their technological history. 
First, a description of LCET innovation dynamics over the past two centuries measured by relative and quantitative accounts of LCET patents and their reliance on science is provided. 
In a subsequent step, the qualitative nature of the scientific and technological knowledge base of LCETs is explored and its evolution over time illustrated. Finally, cross-technology interactions revealed by the network structure of the patent-science citation data are studied. 

\begin{figure}
\centering
\includegraphics[width=.75\linewidth]{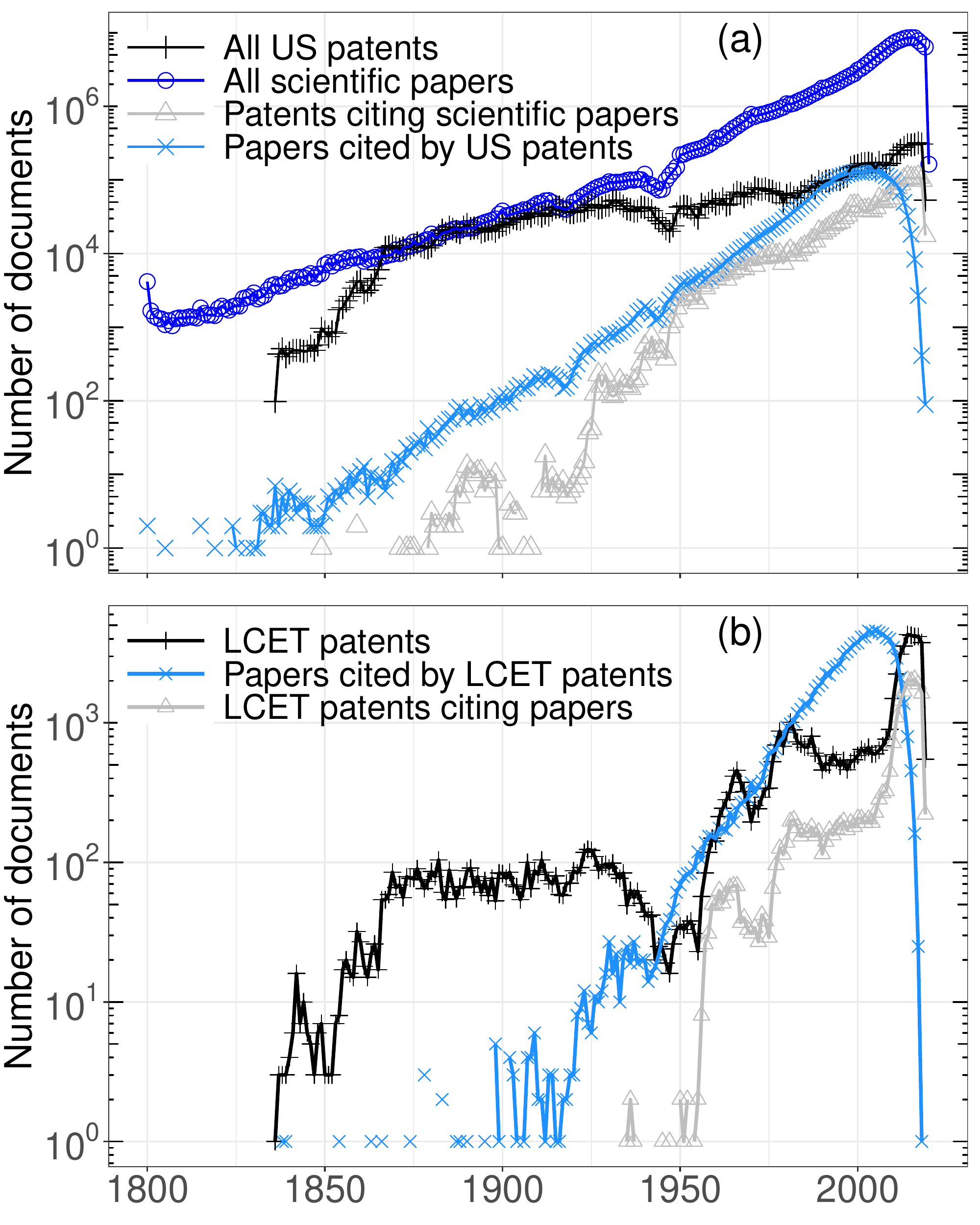}
\caption{Number of patents and scientific articles over time where the year corresponds to the publishing date. (a) Time series for the entire data set. (b) Time series restricted to the LCET subset only.}
\label{fig:full_timeseries}
\end{figure}

\paragraph{Exponential growth in science and patenting}
Fig. \ref{fig:full_timeseries}(a) shows the evolution of the number of patents and papers in the two full data sets, on semi-log axis, suggesting exponential growth in the number of patents and papers, although at different rates. In this figure, the data covers only US patents, but all available papers in the MAG database. The drop during the most recent years can be explained by a (yet) incomplete coverage of most recent papers and patents. A drop in patenting and scientific activity during World War II, and in scientific activity during World War I, is observed.
Fig. \ref{fig:full_timeseries}(a) also indicates exponential growth in the number of papers cited by at least one patent, and the number of patents citing at least one paper.\footnote{Note that this time series will change in the future, as future patents will cite papers that already exist.}

Limiting the data to LCETs, Fig. \ref{fig:full_timeseries}(b) shows the evolution of the number of documents. The fairly large number of LCET patents in the period 1850-1950 is mostly  due to Hydroelectric energy and to some extent Wind, i.e. two technologies that do not tend to cite many papers, as it will be seen. The number of LCET patents citing papers does not start to grow before the second half of the 20th century. 

On average, the time lag between the publication date of a scientific paper and the citations made by patents is 13 years, but it can be much longer. For example, patent US6310406 for Hydro granted in 2001 is referring to Faraday's experimental studies on electricity \citep{faraday1838xiii}. This is not the first time that Faraday's studies were cited, but it indicates that parts of the current scientific knowledge base of LCETs are very old. 

\paragraph{Changes in the relative importance of LCET patents.}
LCET patents account for only a marginal share of the total number of patents (0.1\%-3.3\%), with important fluctuations over time (see Appendix \ref{SI:fig:timeseries_nuc_share}). Broadly speaking, there is a secular decline in the patent share of Renewables until the late 70s when the oil crisis triggered a large and well-documented boom and bust \citep{geels2017socio, grubler2012policies}. This was followed by a renewed surge of interest in the 2000's.  Nuclear energy patents peaked in 1960-70s and accounted for a fairly stable share after the first oil crisis in 1973.

\begin{figure}
	\centering
	\includegraphics[scale=0.8]{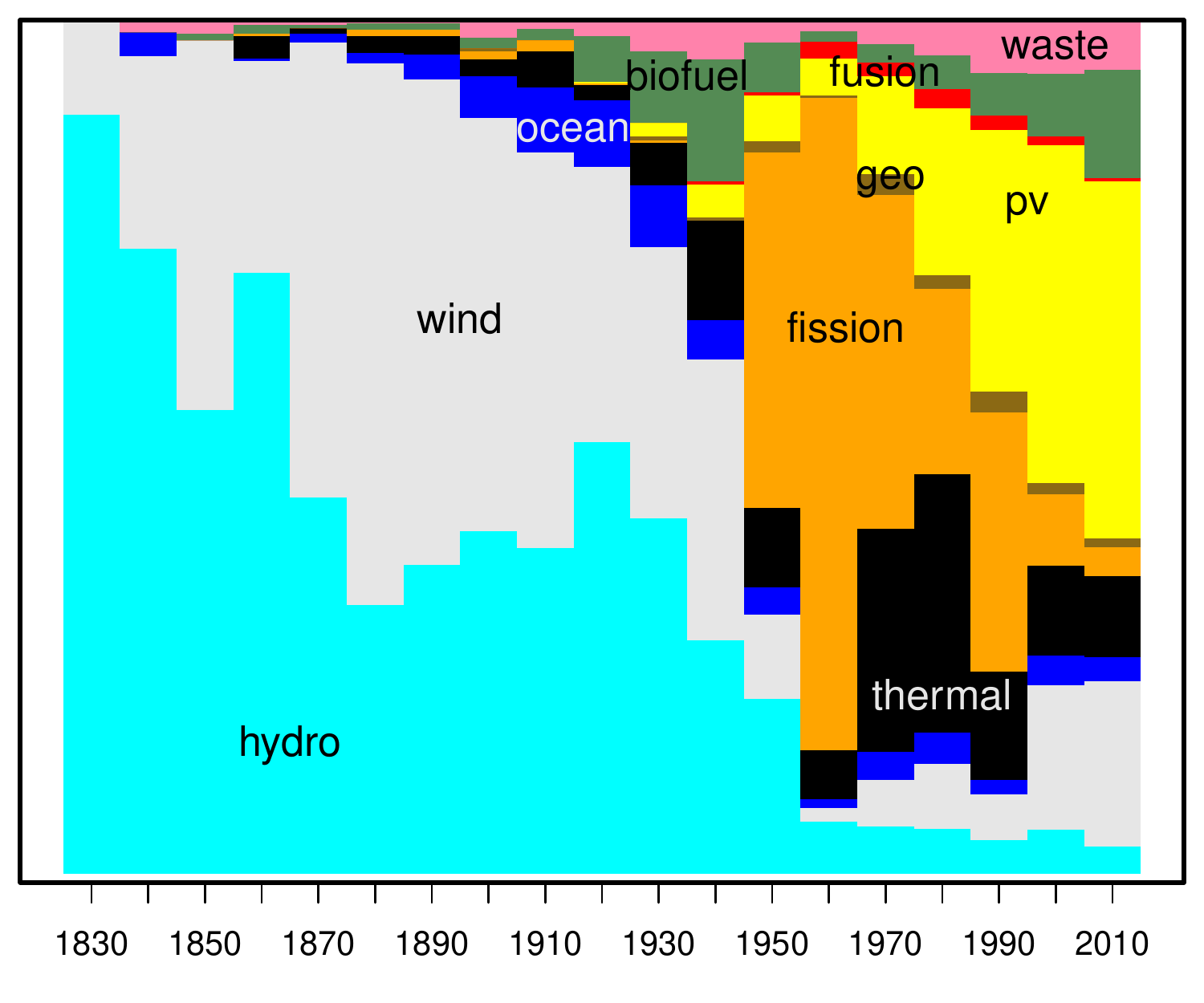}
	\caption{Share of each technology in the total annual number of LCET patents}
	\label{fig:timeseries_shares}
\end{figure}

Fig. \ref{fig:timeseries_shares} shows the evolution of the shares of each technology type within the total of LCET patents. The whole period until 1945 is vastly dominated by Hydro and Wind energy. 
It is known from history that Wind and Hydro are very old technologies, starting with sailing 5,500 years ago \citep{sorensen1991}, watermills for grain milling in the first century BCE \citep{smil2016} and windmills, first documented at the turn of the second millennia in Persia \citep{kenneth1990}. Later, windmills developed in the US alongside railways and the conquest of the west \citep{smil2016}, being used in particular on farms for water pumps. Wind was used in windmills but was not competitive against coal when energy production shifted to electricity at the turn of the 20th century \citep{sorensen1991}. A renewed interest emerged after the oil crisis \citep{kaldellis2011wind}.

The use of Solar ``thermal'' energy is also very old. \citet{perlin2013} documents solar architecture since ancient China, but also development in burning mirrors (from antiquity to the Renaissance) and solar ``hot boxes'' (using the greenhouse effect) in the 18th and 19th century. The first Solar thermal patents in the sample are granted for engines and solar water heaters, as also described by \citet{perlin2013}, but account for only a small share of 19th century patenting. 
Fig. \ref{fig:timeseries_shares} shows that Solar thermal took a fairly constant and significant share of LCET patents throughout the 20th century.
The larger share of Solar thermal patenting in the 1970's-90's, and to some extent now, is not too surprising as the issue of energy storage has always been an important bottleneck for the development of Renewables. In many solar thermal systems, energy is stored in the form of heat.

Biomass has been used since the dawn of civilization, and there have been fairly continuous technological development throughout the last two centuries. Fig. \ref{fig:timeseries_shares} appears consistent with the history of these technologies as reported in \citet{guo2015bioenergy}, who document the very early use of solid biomass (e.g. wood, charcoal), give examples of the use and technical challenges of Biofuels throughout modern history (e.g. for motor vehicles in the 19th century), and describe more recent technological progress, in liquid and gaseous biofuels particularly.

In the data, a striking boom in Nuclear energy technology after World War II is observed. Fission technologies accounted for more than 80\% of all LCET patents in the 1960s. Historians spoke of a ``veritable euphoria'' \citep[pp.452]{cardwell1994}. During this period, the civil use Nuclear energy was strongly pushed by (inter)national initiatives \citep{cowan1990nuclear}.

A steady growth of the relative importance of Solar PV technology is observed. It makes up the highest share from the 1990s onward. 
The potential of the photoelectric effect is known since the late 19th century. But it came with Bell Lab's work on semiconductors in the early 1950's, and the shift from selenium to silicon solar cells, together with a niche market in space exploration, that Solar PV is seen as a major player \citep{perlin2013}. Constant rates of technical progress have contributed to a large diffusion ever since \citep{farmer2016predictable}.

\begin{figure}
	\centering
	\includegraphics[width=\textwidth, trim = {0cm 0cm 0cm 0cm}, clip]{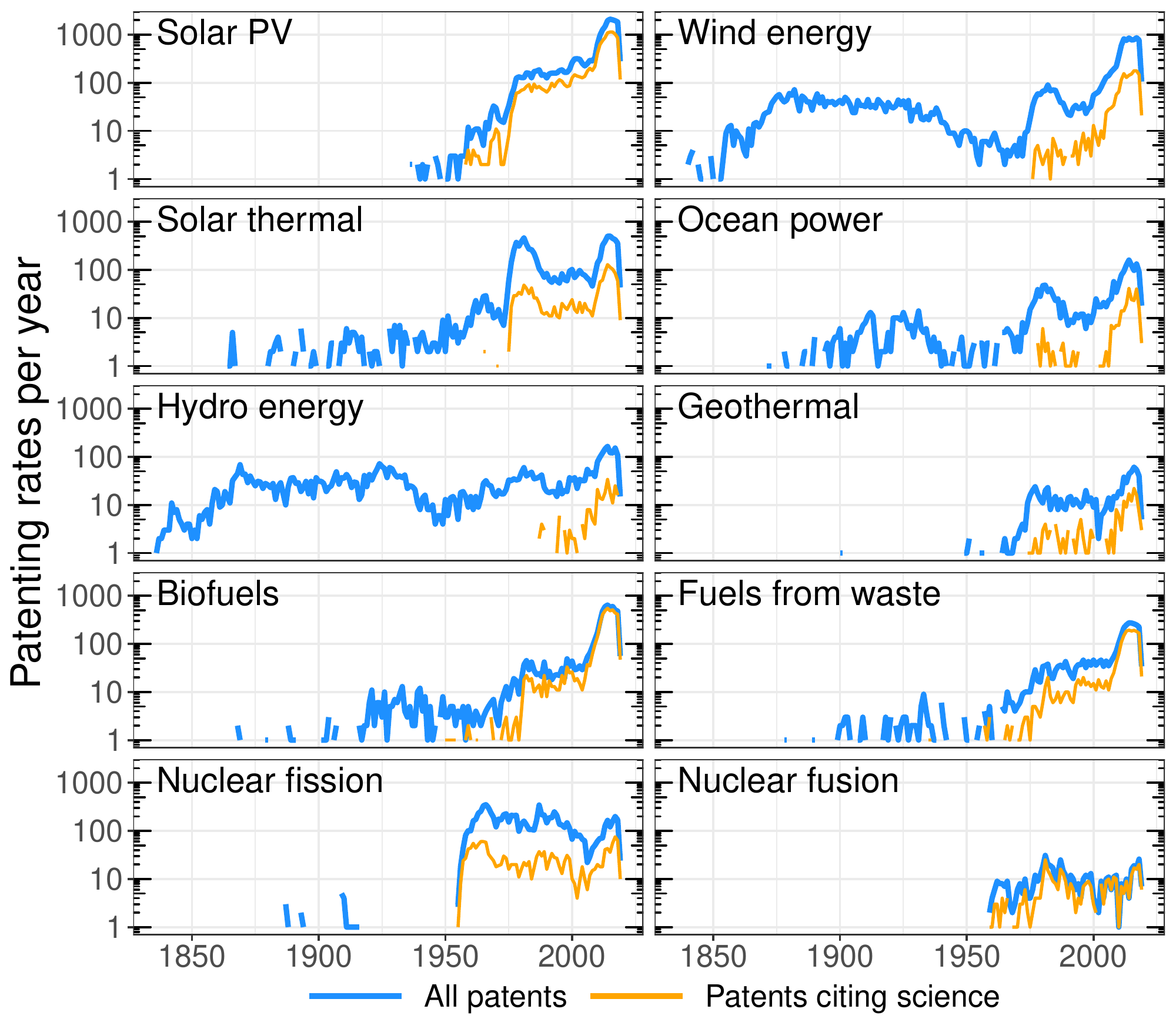}
	\caption{Evolution of the annual number patents in each low-carbon energy category (blue). The thinner line (orange) shows the number of patents that cite at least one paper.}
	\label{fig:green_timeseries_count}
\end{figure}

\paragraph{More and more patents do cite science.}
Fig. \ref{fig:green_timeseries_count} depicts a time series of patent counts per technology type. Hydro and Wind energy are the oldest technologies in the sample. Both grew initially and fluctuated at stable levels around 10-50 patents per year until the World Wars. Ocean power, Solar thermal energy, and fuel-based LCETs began to grow next. These four technologies were not significant until the 1950s with usually less than 10 patent applications per year. 
Fig. \ref{fig:green_timeseries_count} shows that the four ``old'', non-fuel LCETs (Solar Thermal, Wind, Hydro and Ocean) exhibit a boom and bust in patenting activity in the aftermath of the oil crisis and a second rise beginning in the mid 2000s.

\begin{figure}
\centering
  \centering 
    \includegraphics[width=\linewidth, trim = {0cm .8cm 0cm 0cm}, clip]{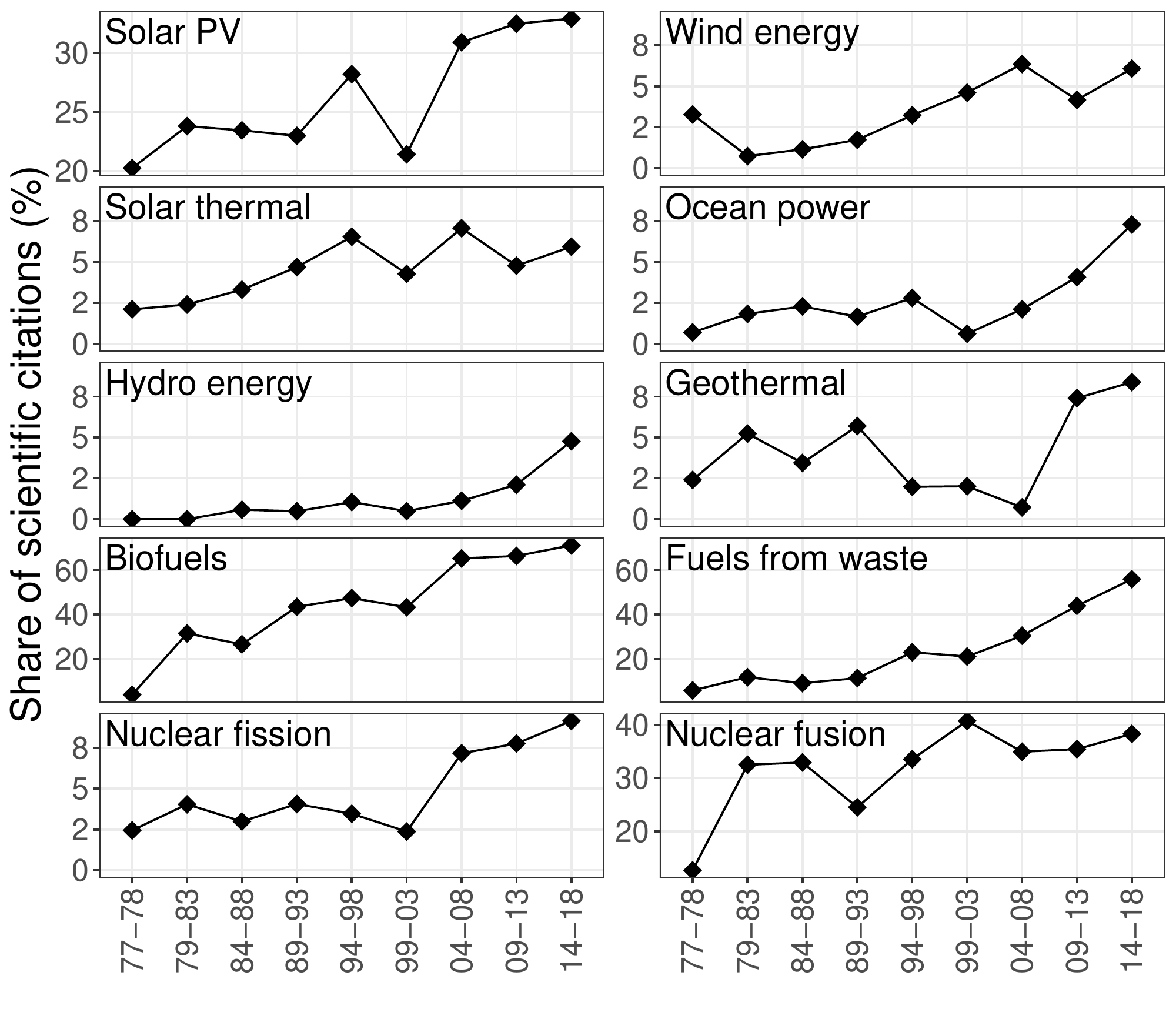}
  \caption{Share of total citations (patents $+$ science) to scientific publications in \%. Citations are aggregated over 4-year time windows. The y-axes of Solar PV, Fusion, Biofuels and Waste are on different scales because they have substantially higher shares of scientific citations. }
    \label{fig:pcs_share}
\end{figure}

Except from few very early Fission patents that may be an artefact of reclassification errors (Sec. \ref{sec:methods_data_pat}), patenting activity in Nuclear technologies grew very quickly postwar, and have remained relatively stable since, with a weak tendency downwards. The number of patent applications in Geothermal energy is modest but relatively stable with around 10-50 patents since the 70s and a possibly increasing trend since 2000.
The rise of Solar PV is exceptional, growing exponentially at high rates from the 1960s onward to become by far the most patented LCET in every single year since 1995. In the years before the highest patenting rates are observed for Nuclear fission which has only the fifth highest patenting rates (out of ten) in the most recent years. Wind energy exhibits a similar but somewhat less extreme increase in patenting than Solar PV since the early 2000s when it became the technological field with the second highest patenting rates.

Fig. \ref{fig:green_timeseries_count} shows the number of patents citing scientific papers (thin orange line). 
It indicates that referencing scientific work is a fairly recent phenomenon and that the share of patents citing non-patent literature is highly heterogeneous among LCETs (Appendix \ref{section:SI_timeaggreg} presents time-aggregated statistics). For example, a large share of PV, Nuclear, Biofuels and Fuels from waste patents refer to science. The share is much lower for Hydro energy and Ocean power. 
The peak of patenting in Wind, Solar thermal, Geothermal and Ocean power after the oil crisis in the 1970s is accompanied by an increase in the relative number of patents citing science.

\paragraph{Rising share of scientific citations in patents.} The rise of science in the LCET knowledge base is also depicted in Fig. \ref{fig:pcs_share}, which shows the share of total citations going to scientific literature (``total'' is defined as the number of patent citations plus number of citations to scientific papers, i.e. excluding other non patent references). Citations in patent documents are used to describe prior art, i.e. to describe the novelty of the invention in relation to pre-existing technological and scientific knowledge embodied in patents and papers \cite[cf.][]{meyer2000does}. 
Note that the scale of the y-axes for PV, Fusion, Biofuels and Waste technologies are different from the others because their share of citations is much higher. 
Biofuels and Fuels from waste have by far the largest shares of scientific citations with more than half of all citations going to scientific papers in recent years. 
This observation can be explained by the fact that these technologies relate mostly to Chemistry, which is known to be science-intensive.
More than 97\% of Biofuels and Waste patents are classified with the Chemistry CPC class ``C''.
The share of scientific citations in these technologies has increased enormously over the last 40 years.
The rise in the share of citations to the scientific literature is also substantial in Solar PV and Nuclear Fusion patents.

Despite lower overall reliance on science, the rise of science can also be observed for the other technologies. Hydro patents cited almost no scientific articles at all in the earlier years, but about 5\% of all Hydro patents did so in the most recent years. Ocean power starts from similar low levels and reaches an even larger share of scientific citations in the last time window. Comparable positive trends can also be observed for Nuclear fission, Solar thermal and Wind energy. Geothermal patents were an exception until the early 2000s, but also appear to experience an increase over the last decade.

Figs. \ref{fig:green_timeseries_count} and \ref{fig:pcs_share}, taken together, make it clear that there is a rise of science in LCETs: more patents do cite science, and a higher share of citations is to scientific articles rather than patents. Could this be an artefact of the data? Perhaps early patents contain citations that are not detected, or citation practices were different in the past? For instance, it is likely that patent-to-paper citations are more likely to be missed in early rather than current patents, and it is likely that early patents were citing books and textbooks more than papers. These issues are explored in Appendix \ref{sec:si_cleaning_early_pats} where it is argued that the rise of science is not an artefact of the data. First of all, checking many early patents manually confirmed that they did not contain many references to scientific materials. Second, reading the early patents and comparing them with more recent patents qualitatively confirms what is known from historians \citep{mokyr1990twenty,freeman1997economics,headrick2009technology}: the 19th century was marked by individual inventors, tinkering designs by trial and error, and making relatively few references to formal science. The turn of the century was associated with the emergence of larger R\&D labs, with Edison's Menlo labs being the well-known example for the case of electricity. After World War II, the `linear model' (science drives technology, which drives the economy) dominated and resulted in more and more science-driven, technology-push advances. Later in the century, science and technology became more inter-related, with a rise in academic patenting, rise of industrial scientific contributions (Bell labs, IBM, etc.), rise of the number of PhDs hired in industry, etc. leading to a further increase in the degree to which patents cite state-of-the-art scientific results.

\subsection{Evolution of the low-carbon energy knowledge base}

We now investigate the \emph{composition} of the scientific knowledge that is cited by each technology type. Because there are too few patents citing papers in earlier years and still a limited number now, we focus on recent years and aggregate periods. To strike the right balance between a high enough number of periods and a high enough number of patents within each period, three periods are shown: 1976-1990, 1991-2005, and 2006-2019. For each technology, the share of the four most important scientific domains cited (see Sec. \ref{sec:methods_networks}) are presented. Scientific domains are classified by Web of Science (WoS) categories. A more detailed analysis can be found in Appendix, Fig. \ref{fig:SI_science_reliance_change}, Sec. \ref{sec:SI_networks}.

\begin{figure}
	\centering
	\includegraphics[scale=1]{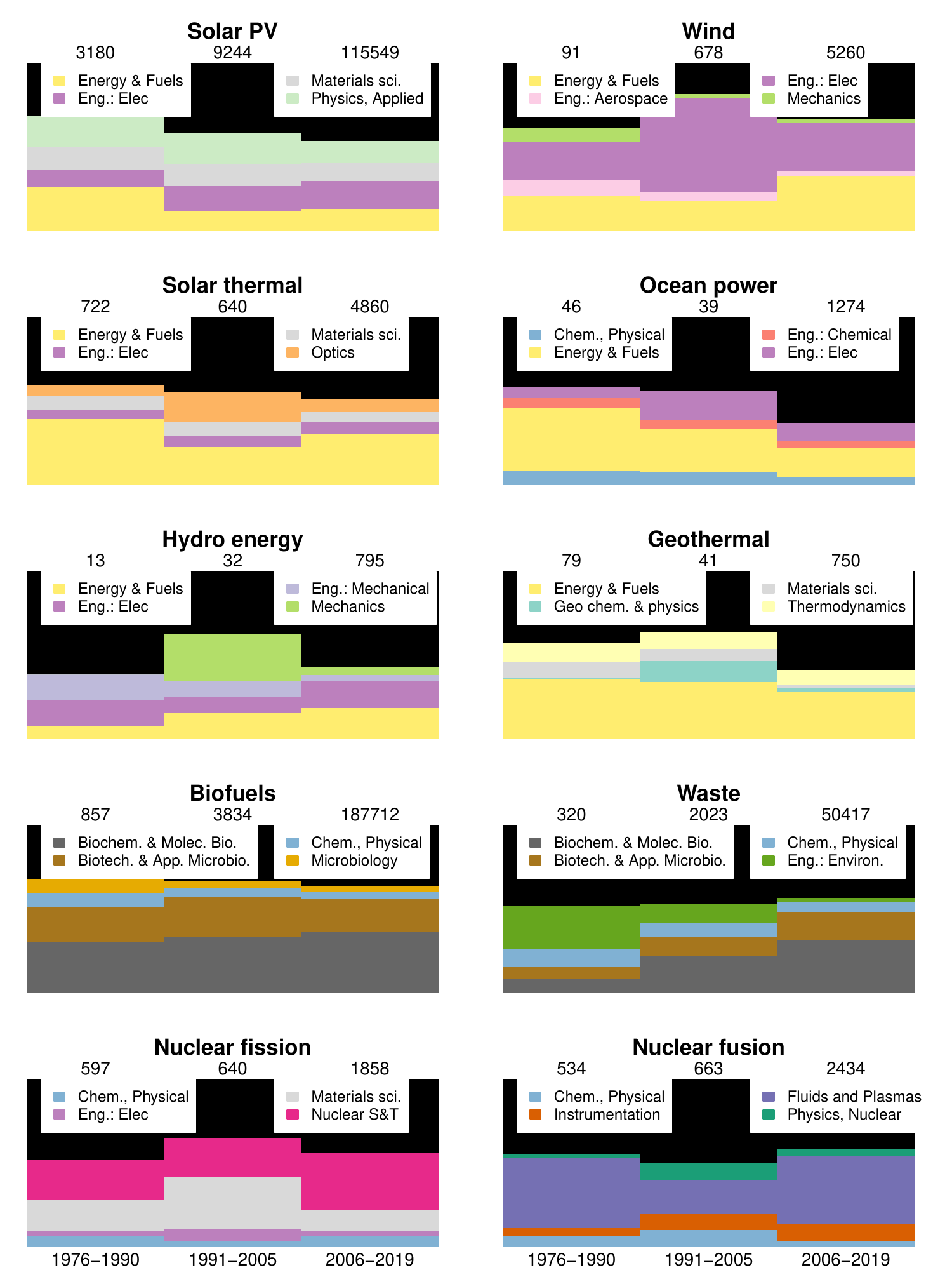}
	\caption{Prevalence of scientific domains in the knowledge of each LCET, for three periods (1976-90, 1991-2005, 2006-19). Colors are consistent across charts, except black which is ``all others''. The numbers on the top horizontal axis are the total numbers of citations made to papers during the period.}
	\label{fig:green_reliance_on_wos}
\end{figure}
\FloatBarrier

Three key findings emerge. First, as expected, the field ``Energy \& Fuels'' is well represented in all technologies, except for the two Nuclear and the two Fuels. Second, in each technology domain-specific science is found: ``Material Science'' for Solar PV, ``Aerospace engineering'' for Wind, ``Optics'' for Solar thermal, ``Chemical Engineering'' for Ocean, ``Mechanics/mechanical engineering'' for Hydro, ``Geology'' and ``Thermodynamics'' for Geothermal, 
``Biochemistry'' for Biofuels, ``Environmental Engineering'' for Waste, ``Nuclear Science and Technology'' for Fission, and ``Fluids and Plasma'' for Fusion. Biofuels and Fuels from waste rely most strongly on Biochemistry and Microbiology. Third, in all technologies except the two Fuels, Geothermal and Fusion, a significant share of ``Electrical Engineering'' is seen.

Fields of scientific research can be grouped into applied and basic science. Applied sciences are more closely related to the technical application of knowledge \citep{OECD2015}. 
Following the heuristic identification rule used by \citep{persoon2020science}, applied sciences are identified by their membership to the WoS aggregate category that is named ``Technology''.\footnote{
This category contains 21 more finely grained scientific fields, i.e. Acoustics, Automation \& Control Systems, Computer Science, Construction \& Building Technology, Energy \& Fuels, Engineering, Imaging Science \& Photographic Technology, Information Science \& Library Science, Instruments \& Instrumentation, Materials Science, Mechanics, Metallurgy \& Metallurgical Engineering, Microscopy, Nuclear Science \& Technology, Operations Research \& Management Science, Remote Sensing, Robotics, Science \& Technology Other Topics, Spectroscopy, Telecommunications, Transportation. See also  \url{https://images.webofknowledge.com/images/help/WOS/hp_research_areas_easca.html}
} 
One tendency becomes apparent: Science-intensive technologies tend to rely to a larger extent on basic, rather than applied science. For example, science-intensive technologies tend to draw more intensively from disciplines of basic research, e.g. biology and biochemistry for non-fossil fuels and physics for PV and Fusion. In contrast, Wind and Hydro largely rely on engineering and electronics which are classified as fields of applied research. This tendency is even more pronounced in the ranking of the top-10 WoS fields shown in the Fig. \ref{fig:SI_science_reliance_change} in the Appendix.

Fig. \ref{fig:green_reliance_on_wos} also shows that there is high stability in the composition of the knowledge base of each technology, the main exception being perhaps the increasing share of Biochemistry and Molecular Biology in Fuels from Waste patents. 

Overall, the results on the composition of the knowledge base of each technology (Fig. \ref{fig:green_reliance_on_wos}) and the differences in science intensity between technologies documented in Figs. \ref{fig:green_timeseries_count} and \ref{fig:pcs_share} broadly confirm historical knowledge and intuition, but also reveal less well appreciated facts (for instance the relatively low share of scientific citations in Nuclear Fission, despite Nuclear Energy being based on fundamental physics).

\FloatBarrier

\subsection{Evolution of technology-science networks}
From a policy perspective, it is important to know whether technologies rely on similar scientific knowledge bases. For example, if two technologies heavily rely on the same scientific field, scientific progress in this field would likely benefit both technologies in terms of knowledge spillovers. If a scientific field is relevant for many technology types, targeted research investments could simultaneously foster innovation activities across technologies.

To evaluate the similarity of two technologies in terms of their scientific knowledge base, we compute the cosine similarities of technology pairs, i.e. the relative shares of citations distributed over scientific fields are compared. If two technologies cite scientific fields in the same proportions, a cosine similarity equal to one is obtained. If two technologies cite completely disjoint sets of scientific fields, the cosine similarity is equal to zero (see Sec. \ref{sec:methods_networks} for details).

\begin{figure}
	\centering
	\includegraphics[width = .8\textwidth, trim = {15cm .5cm 0.3cm .5cm}, clip]{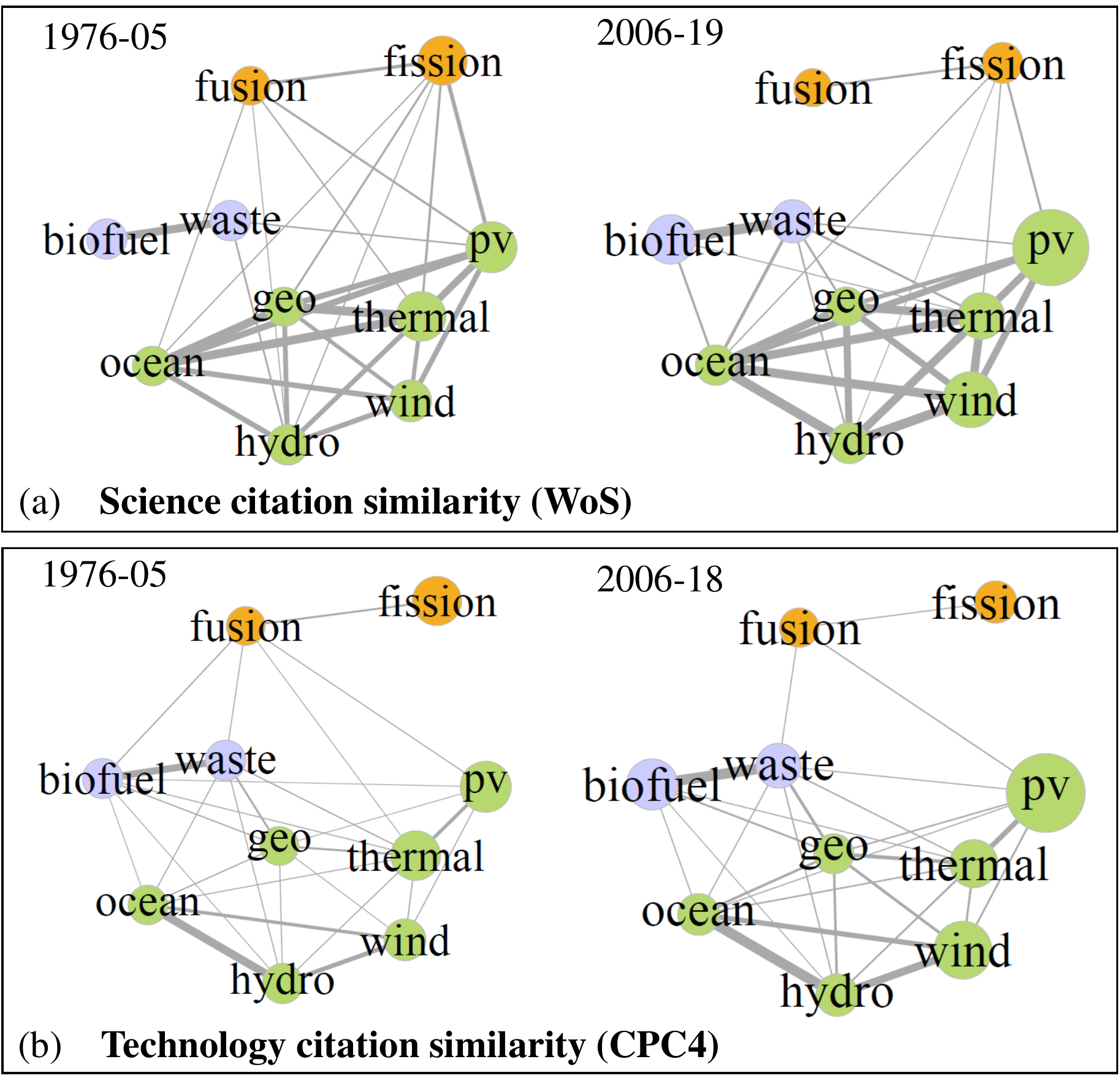}
	\caption{Cosine similarity networks based on shares of (a) citations to scientific fields (WoS) and (b) citations to CPC 4-digit technology classes. A link between a pair of LCETs indicates the cosine similarity of their references to scientific fields and technology classes, respectively. For clarity only the largest two thirds of links are shown.}
	\label{fig:similarity}
\end{figure}

Fig. \ref{fig:similarity}(a) visualizes the cosine similarities of scientific citations by LCETs as networks for two different time periods. Edges are thicker if the cosine similarity is higher. The visualization shows three broader clusters: Nuclear energy technologies, Biofuels and Fuels from waste and the six non-fuel based renewable technologies.
Renewables form a densely connected cluster, indicating that they are relying on similar scientific fields. 
This relationship even gets stronger over time with the average pairwise similarity for Renewables increasing by 22\% from 0.63 to 0.76. 
The driving forces behind this development are Wind and Hydro energy which became scientifically more similar to all other Renewables.

The similarity in scientific reliance for Biofuels and Waste is even more pronounced than for Renewables. 
The cosine similarity lies at 0.84 before 2005 and increases to almost one for the latter period. Since 2006, 50\% of Waste patents have been classified as Biofuels and around a quarter of Biofuels patents have been classified as Waste technologies. While still large, the frequent co-classifications of the two technology types are insufficient to explain the extremely large similarity values for scientific citations.
Appendix \ref{sec:SI_robust} shows that the results are very robust against potential impinging factors like citation confidence scores and patent co-classifications.

In contrast to the other clusters, the similarity of scientific citations between the two nuclear power technologies has decreased over time. Remarkably, every single link between Nuclear and any other technology has become weaker over time with only a single exception (the link between Fission and Ocean remained essentially constant).
More details on the underlying scientific fields and alternative, more fine-grained bibliographic coupling networks are provided in Appendix \ref{sec:SI_networks}-\ref{sec:SI_bibcoup}. 

Fig. \ref{fig:similarity}(b) repeats the analysis but when using patent-to-patent instead of patent-to-papers citations.
Here, a link indicates whether LCETs cite similar technological classes (instead of scientific fields).
A similar separation between Nuclear energy, Renewables and Fuels becomes evident. Furthermore, the time trends observed above are confirmed: the average link size increases between Renewables (+70\% increase of average cosine similarity) and between Biofuels and Waste (+40\%) but decreases between Fusion and Fission (-75\%).
Besides the Biofuels-Waste link, there is also exceptionally strong interconnections between the triangle Wind-Hydro-Ocean which has significantly increased in time. While the relationship between Geothermal and PV is less pronounced, both technologies show strong ties to Solar thermal.
The three clusters of LCETs revealed by the network analysis (Renewables, Fuels and Nuclear technologies) are in line with intuition. However, there is substantial differences in how tightly nodes are connected in these clusters and how clusters are interlinked. These relationships are not fixed but can change in time.

\FloatBarrier

\section{Discussion and conclusions}
\label{conclusion}

This paper offers an analysis of two centuries of the technological history of LCETs using an almost complete data set on LCET patents in the US and scientific papers cited by them. 
This is the most comprehensive and detailed analysis describing the evolving knowledge base of ten technologies responsible for all low-carbon electricity generation, covering 200 years of technological and scientific history of low-carbon energy. This approach became possible only very recently due to previous limitations in the data availability, quality and processing power. 

Four main results emerge.

First, patenting activity in LCET is old, but has fluctuated widely. LCET patenting used to be dominated by Hydro and Wind, until the massive rise in importance of Fission during the first few decades post World War II. In recent decades, the most striking is the rise of Solar PV and the comeback of Wind. Ocean power, Geothermal and Fusion are fairly minor, while Solar thermal, Biofuels and Fuels from waste have remained secondary but non-negligible throughout the 20th century.

Second, overall there is a clear increasing importance of science in LCET patents. Specifically, there is a strong increase in both the share of LCET patents citing papers and in the share of citations that are to papers rather than to patents.

Third, this increasing trend is consistent across different types of LCETs, but striking and large differences remain. Biofuels and Fuels from waste, Solar PV and Nuclear fusion are strongly science-intensive, while Hydro and Geothermal are much less so. Moreover, LCETs with higher shares of citation to the scientific literature tend to draw more on basic rather than applied sciences.

Fourth and finally, we find that while each LCET relies on technology-specific scientific fields (e.g. optics for solar thermal), there are also significant overlaps in the knowledge bases on several LCETs, for instance Electrical Engineering for Wind, PV, Hydro, Geothermal and Ocean Power, or Biochemistry and Molecular Biology for Biofuels and Fuels from Waste.

Before discussing the practical value of these results, some limitations need to be mentioned. First of all, not all technologies are patented and patenting propensity may vary across fields. Second, there may have been a change in citation culture over time, with an increasing propensity to report citations to science. Third, the data is only about US patents. Although it is known that the US represents the technological frontier for green energy technologies \citep{albino2014understanding}, it would be interesting to compare with other countries for the periods where this is possible. Fourth, there remains some issues with data quality, with some documents and citations missing, and imperfect matching between patents and papers. Fifth, the research heavily relies on classification systems, which are known to be imperfect and changing over time \citep{lafond2019long}.
More generally, the conceptual framework assimilates patents to technology and applied development, and papers to science and fundamental knowledge, but of course the boundary between the two is not clear and may be evolving over time and across fields.

This paper is limited to a descriptive analysis of LCETs, but this is an essential first step. It became possible only recently for reasons of data availability and data processing power. This analysis can lay the foundations for deeper research into the causes and consequences of heterogeneity in technological development. One avenue for further research would be to explore if the relationship between patent value and research quality found in Refs. \cite{watzinger2019standing,poege2019science} holds within each LCETs, and for LCETs as a group. A second interesting extension would be to compare LCETs with other groups, such as non-green energy technologies or non energy-related technological domains. A third avenue for research would be to link this data to direct and objective measures of technological progress, such as cost and technological features. It is known that costs tend to decrease exponentially at a technology-specific rate \citep{farmer2016predictable}, and that this rate correlates with centrality of a technology in the patent citation network \citep{triulzi2020estimating}. But it is difficult to establish whether the key underlying driver of technological progress is scientific, technological, linked to industrial manufacturing development, or external \citep{witajewski2015bending}. Obtaining data on R\&D for detailed technological domains is notoriously difficult, but would eventually be needed for understanding how responsive each technology is to various policy support schemes. Finally, and relatedly, there are several opportunities for a better understanding of cause and effects relationships, using relatively `exogenous' events such as the oil crisis or nuclear accidents.

In spite of these limitations, the results of this analysis are informative for policymakers and R\&D managers who are challenged by two major questions when developing strategies to foster LCET development and diffusion. First, which technology or combination of technologies should be chosen? And second, how to design effective support instruments? 
The analysis does not provide ultimate answers, but draws a comprehensive picture of the scientific fields underlying each LCETs, from which a number of preliminary policy implications can be derived. 
First, LCETs that are more science-intensive (e.g. Solar PV, Nuclear Fusion, Biofuels) are likely to need significant continued investment in scientific research to keep making progress, while other less science-intensive technologies (e.g. Hydro and Wind) are expected to benefit more from applied development and engineering.
Second, there are three fairly separate clusters of technologies (Nuclear, Fuels, and Renewables). This suggests that investing in science underlying a specific LCET is likely to be beneficial for other LCETs in the same cluster, but less for LCETs belonging to other clusters. If spillovers are sufficiently strong to generate increasing returns, the benefits of technological specialization are higher if R\&D resources are focused on one cluster instead of being spread across technologies belonging to different clusters.

\FloatBarrier

\section*{Data availability}
All data used for this publication is available for download in a separate data publication registered under  \url{doi.org/10.4119/unibi/2941555}. It can be used and modified in agreement with a \href{https://creativecommons.org/licenses/by/4.0/legalcode}{CC BY 4.0} license. 

\section*{Acknowledgements}
This paper was written while Kerstin H{\"o}tte was visiting INET Oxford. Anton Pichler and Fran\c{c}ois Lafond would like to thank Partners for a New Economy, the Oxford Martin Programme on the Post-Carbon Transition, Baillie Gifford, and the Institute for New Economic Thinking at the Oxford Martin School. Kerstin H{\"o}tte wants to thank the Bielefeld Graduate School for Theoretical Sciences, the Bielefeld Graduate School for Economics and Management and the National Academic Foundation. We would like to thank Peter Persoon for comments and for sharing a pre-print of his work with us. Further thanks to Herbert Dawid and J. Doyne Farmer for helpful feedback. Many thanks to Matthew Marx and his colleagues from the Reliance on Science project for practical help and the provision of data. 
\FloatBarrier


\bibliographystyle{unsrtnat}

\bibliography{refs.bib}

\FloatBarrier
	\newpage
\appendix
\section*{Appendix}
\label{SI}

\section{Aggregate statistics of LCETs and citations to papers.} 
\label{section:SI_timeaggreg}

Table \ref{t:citestats} aggregates key citation statistics over the whole time horizon. It is seen that the scientific dependence of LCETs is highly heterogeneous itself, as already described by \citet{persoon2020science}. For example, roughly 75\% of Biofuels patents, 66\% of Nuclear fusion and more than half of Solar PV and Fuels from waste patents cite scientific research. All other technologies cite scientific articles substantially less frequently ranging between 21-4\%. Hydroelectric patents are least science-intensive with 4\% but with a increasing trend in recent time. 

The total numbers of citations to science per technology class is similarly diverse, ranging from 846 Geothermal citations to more than 190,000 citations for Biofuels patents and almost 130,000 citations for Solar PV. The diversity of scientific citations is also reflected in the statistics of citations per patent. 
Interestingly, we find more citations per patent for Solar PV than for fusion when looking only at patents which cite at least on scientific paper. In contrast, and despite its relatively high reliance on science, Fission patents which cite papers cite on average only 2.4 papers.
Biofuels and Fuels from waste are again outstanding with 43.7 and 27.9 citations per patent. 

\begin{table}[b]
\centering
\resizebox{\textwidth}{!}{
\begin{tabular}{lrrrrrr}
  \hline
Energy technology & Total patents & Citing patents & Citing/total patents & Scientific citations & Citations per patents (all) & Citations per patents (citing) \\ 
  \hline
Solar PV & 21,237 & 11,541 & 0.54 & 128,079 & 6.03 & 11.10 \\ 
  Wind energy & 11,973 & 1,590 & 0.13 & 6,030 & 0.50 & 3.79 \\ 
  Solar thermal & 8,706 & 1,460 & 0.17 & 6,236 & 0.72 & 4.27 \\ 
  Ocean power & 2,090 & 271 & 0.13 & 1,360 & 0.65 & 5.02 \\ 
  Hydroelectric & 5,419 & 229 & 0.04 & 840 & 0.16 & 3.67 \\ 
  Geothermal & 846 & 177 & 0.21 & 874 & 1.03 & 4.94 \\ 
  Biofuels & 5,868 & 4,401 & 0.75 & 192,437 & 32.79 & 43.73 \\ 
  Fuels from waste & 3,422 & 1,891 & 0.55 & 52,797 & 15.43 & 27.92 \\ 
  Nuclear fission & 9,507 & 1,730 & 0.18 & 4,159 & 0.44 & 2.40 \\ 
  Nuclear fusion & 596 & 391 & 0.66 & 3,692 & 6.19 & 9.44 \\ 
  \hline
  Overall & 65,305 & 22,017 & 0.34 & 396,504 & 6.39 & 11.63 \\ 
   \hline
\end{tabular}
}
\caption{Aggregate statistics of low-carbon energy technologies and citations to scientific literature. The last row \textit{overall} corresponds to total numbers for columns \textit{total patents}, \textit{citing patents} and \textit{scientific citations} and for averages for the other columns.
}
\label{t:citestats}
\end{table}
\FloatBarrier

\section{Statistics of the science-reliant LCET data subset}
\label{si:stats_on_subset}
The following tables give an overview of the subset of LCET patents and papers we analyze here. Table \ref{tab:gen_stats} shows the counts of patents, papers and citations for all LCETs and per LCET type and the number of citation links with highest and lowest confidence score (CS). The confidence score indicates the reliability of the citation link inferred by the stochastic paper-patent matching procedure by \citet{marx2019reliance}. The average CS of our data ranges between 7.41 and 9.34 which indicates an expected precision rate of more than 99.7\%. 

Patent citations can be added to the document by the examiner or applicant. In the data, 72-96\% of citations to science are added by the applicant and between 1-10\% by the examiner. The type of the remaining fraction is unknown. 
13-63\% of citations are retrieved from the text body only and 30-79\% from the front page only. The remaining citations are made in both, the text body and the front page. 

Table \ref{tab:time_coverage_data} shows the age characteristics of the patents and papers. 
In Table \ref{tab:papers_most_cited}, information on the most cited paper per technology group is shown. The title and the DOI, if available, can be used to search manually for the paper online. As a validation test, the papers without DOI/Journal were checked manually.\footnote{
Wind: \url{https://docs.lib.purdue.edu/icec/1508}. 
Solar thermal: \url{https://www.osti.gov/biblio/7338428}. 
Fuels from Waste: ISBN 9781560325536} 
These are a technical report and a conference contribution which have not been published in an official conference-proceeding or journal and two books.

\begin{landscape}
\begin{table}[ht]
\centering \footnotesize
\begin{tabular}{p{2.25cm}p{1.5cm}p{1.5cm}p{1.5cm}p{1.5cm}p{1.5cm}p{1.5cm}p{1.5cm}p{1.5cm}p{1.5cm}p{1.5cm}}
  \hline
& \# patents & \# papers & \# citations & \# CS = 10 & \# CS = 3 & Avg CS & \% app & \% exm & \% text & \% front \\ 
  \hline
Solar PV & 11,541 & 43,015 & 128,079 & 77,370 & 7,064 & 8.85 & 90.61 & 2.96 & 21.64 & 70.82 \\ 
  Wind & 1,590 & 3,441 & 6,030 & 3,873 & 275 & 8.60 & 81.33 & 6.63 & 14.64 & 80.17 \\ 
  Solar thermal & 1,460 & 3,696 & 6,236 & 3,871 & 459 & 8.74 & 86.79 & 2.74 & 27.39 & 68.14 \\ 
  Energy from sea & 271 & 836 & 1,360 & 818 &  24 & 9.03 & 96.03 & 1.69 & 25.15 & 65.22 \\ 
  Hydroelectric & 229 & 637 & 840 & 558 &  26 & 9.13 & 95.36 & 2.02 & 12.86 & 70.95 \\ 
  Geothermal & 177 & 560 & 874 & 571 &  59 & 8.82 & 91.88 & 0.69 & 22.88 & 73.34 \\ 
  Biofuels & 4,401 & 44,941 & 192,437 & 69,394 & 34,240 & 7.41 & 98.12 & 0.96 & 62.84 & 29.97 \\ 
  Fuels from waste & 1,891 & 18,503 & 52,797 & 26,493 & 5,296 & 8.11 & 96.76 & 0.88 & 45.31 & 50.12 \\ 
  Nuclear fission & 1,730 & 2,576 & 4,159 & 1,807 & 350 & 8.07 & 72.71 & 9.55 & 41.21 & 53.74 \\ 
  Nuclear fusion & 391 & 1,651 & 3,692 & 2,940 & 140 & 9.34 & 78.90 & 2.17 & 18.88 & 74.65 \\ 
  \hline
  All & 22,017 & 103,645 & 396,504 & 187,695 & 47,933 & 8.04 & 94.61 & 1.81 & 44.95 & 48.18 \\ 
  \hline
\end{tabular}
\caption{Overview statistics of subset of green patents citing scientific papers. Columns show the number of (1) unique patents, (2) unique papers, (3) citation links, (4) citations with highest confidence score (CS = 10), (5) lowest confidence score (CS = 3), (6) the average confidence score, (7) the share of applicant and (7) examiner added citations (remaining share is of unknown type), (8) share of citations made exclusively in the text body or (9) front page of the patent document (remaining shares account for citations made in both).} 
\label{tab:gen_stats}
\vspace{1cm}

\centering \footnotesize
\begin{tabular}{p{2.25cm}p{2.1cm}p{2.1cm}p{2.1cm}p{2.1cm}p{2.1cm}p{2.1cm}p{2.1cm}}
  \hline
 & Oldest patent & Youngest patent & Avg year patent & Oldest paper & Youngest paper & Avg year paper & Avg lag \\ 
  \hline
Solar PV & 1936 & 2019 & 2012.02 & 1863 & 2018 & 1999.30 & 12.72 \\ 
  Wind & 1960 & 2019 & 2012.16 & 1878 & 2016 & 1999.27 & 12.89 \\ 
  Solar thermal & 1961 & 2019 & 2008.63 & 1874 & 2016 & 1994.72 & 13.91 \\ 
  Energy from sea & 1937 & 2019 & 2013.17 & 1923 & 2016 & 1998.88 & 14.29 \\ 
  Hydroelectric & 1987 & 2018 & 2013.39 & 1838 & 2016 & 1993.70 & 19.69 \\ 
  Geothermal & 1967 & 2019 & 2010.51 & 1949 & 2015 & 1994.63 & 15.88 \\ 
  Biofuels & 1945 & 2019 & 2013.94 & 1854 & 2017 & 1995.47 & 18.48 \\ 
  Fuels from waste & 1935 & 2019 & 2013.66 & 1898 & 2017 & 1996.40 & 17.26 \\ 
  Nuclear fission & 1950 & 2019 & 1994.66 & 1902 & 2016 & 1980.62 & 14.05 \\ 
  Nuclear fusion & 1959 & 2019 & 2006.09 & 1883 & 2015 & 1987.62 & 18.47 \\ 
  \hline
  All & 1935 & 2019 & 2012.89 & 1838 & 2018 & 1996.65 & 16.23 \\ 
  \hline
\end{tabular}
\caption{Age characteristics of papers and LCET patents which cite scientific papers. Columns show (1) the grant year of the oldest patent, (2) average age of patents, (3) publishing year of oldest cited paper, (4) average age of cited papers and (7) average citation lag.} 
\label{tab:time_coverage_data}
\end{table}

\begin{table}[ht]
\centering \scriptsize
\begin{tabular}{p{2cm}p{5cm}p{0.25cm}p{2cm}p{4cm}p{2.75cm}p{0.25cm}p{1cm}p{1cm}}
  \hline
 & Paper title & Year & WoS field & DOI & Journal/Conference name &  & \# cit. (techn) & \# cit. (total) \\ 
  \hline
Solar PV & A low cost high efficiency solar cell based on dye sensitized colloidal tio2 films & 1991 & Chem., Multidis. & 10.1038/353737a0 & Nature & (J) & 224 & 225 \\ 
  Wind & Wind energy handbook & 2001 & Energy \& Fuels &  &  &  &  29 &  29 \\ 
  Solar thermal & Thermochemical processes for the production of hydrogen from water & 1976 & Chem., Physical &  &  &  &  19 &  42 \\ 
  Energy from sea & Application of solar and wave energies for long range autonomous underwater vehicles & 2002 & Energy \& Fuels & 10.1163/156855302317413736 & Advanced Robotics & (J) &  12 &  13 \\ 
  Hydroelectric & Literature review efficacy of various disinfectants against legionella in water systems & 2002 & Biotechn. \& Appl. Microbiol.& 10.1016/S0043-1354(02)00188-4 & Water Research & (J) &  16 &  16 \\ 
  Geothermal & Experimentation and modelling of an innovative geothermal desalination unit & 1999 & Eng., Chem. & 10.1016/S0011-9164(99)00133-2 & Desalination & (J) &  10 &  28 \\ 
  Biofuels & A general method applicable to the search for similarities in the amino acid sequence of two proteins & 1970 & Biochem. Research Methods & 10.1016/0022-2836(70)90057-4 & Journal of Molecular Biology & (J) & 631 & 715 \\ 
  Fuels from waste & Handbook on bioethanol production and utilization & 1996 & Energy \& Fuels &  &  &  & 104 & 409 \\ 
  Nuclear fission & Fission of elements from pt to bi by high energy neutrons & 1948 & Physics, Nuclear & 10.1103/PhysRev.73.1135 & Physical Review & (J) &  71 &  71 \\ 
  Nuclear fusion & Fundamental limitations on plasma fusion systems not in thermodynamic equilibrium & 1997 & Physics, Multidis. & 10.1063/1.872556 & Physics of Plasmas & (J) &  38 &  38 \\ 
  \hline
  All & A general method applicable to the search for similarities in the amino acid sequence of two proteins & 1970 & Biochem. Research Methods & 10.1016/0022-2836(70)90057-4 & Journal of Molecular Biology & (J) & 715 & 715 \\ 
   \hline
\end{tabular}
\caption{Most cited papers by all technologies and by technology type. Column (7) and (8) shows the number of citations received within the technology group and from all technologies, respectively.} 
\label{tab:papers_most_cited}
\end{table}

\begin{table}[ht]
\centering \footnotesize
\begin{tabular}{p{2.25cm}p{1cm}p{13cm}p{2cm}p{0.5cm}p{0.75cm}}
  \hline \footnotesize
 & Number & Title & CPC code & Year & \# cites \\ 
  \hline
Solar PV & 10020449 & Composition for anode buffer layer of organic thin film solar cell and organic thin film solar cell & Y02E10/549 & 2018 & 662 \\ 
  Wind & 7075189 & Offshore wind turbine with multiple wind rotors and floating system & Y02E10/723; Y02E10/727 & 2006 & 102 \\ 
  Solar thermal & 8187549 & Chemical reactors with annularly positioned delivery and removal devices, and associated systems and methods & Y02E10/41 & 2012 & 114 \\ 
  Energy from sea & 10041466 & Wave-powered devices configured for nesting & Y02E10/38 & 2018 &  79 \\ 
  Hydroelectric & 6927501 & Self-powered miniature liquid treatment system & Y02E10/223; Y02E10/28 & 2005 & 157 \\ 
  Geothermal & 7492053 & System and method for creating a networked vehicle infrastructure distribution platform of small wind gathering devices & Y02E10/12 & 2009 &  92 \\ 
  Biofuels & 10006012 & Polypeptides having cellulolytic enhancing activity and polynucleotides encoding same & Y02E50/16 & 2018 & 1,590 \\ 
  Fuels from waste & 7364890 & Thermal tolerant avicelase from Acidothermus cellulolyticus & Y02E50/343 & 2008 & 1,422 \\ 
  Nuclear fission & 2714577 & Neutronic reactor & Y02E30/40 & 1955 &  58 \\ 
  Nuclear fusion & 10049774 & Systems and methods for forming and maintaining a high performance FRC & Y02E30/122; Y02E30/16 & 2018 &  69 \\ 
  \hline
  All & 10006012 & Polypeptides having cellulolytic enhancing activity and polynucleotides encoding same & Y02E50/16 & 2018 & 1,590 \\ 
   \hline
\end{tabular}
\caption{Patent with the highest number of citations made to scientific articles, for each technology. 
} 
\label{tab:patents_most_citing}
\end{table}

\end{landscape}

In Table \ref{tab:patents_most_citing}, we show the patents per technology group with the highest number of citations to science. Patents and the citation links made by the patent can be manually checked on the website of USPTO using the patent search by number.\footnote{Patent search by number: \url{http://patft.uspto.gov/netahtml/PTO/srchnum.htm}} 
Table \ref{tab:fiels_most_cited} gives an overview of the most cited fields of research and Table \ref{tab:journal_most_cited} summarizes the most cited journals and conference proceedings. 
Table \ref{t:greentech_details} shows relevant statistics on the most granular classification level possible.

\begin{table}[ht]
\centering \footnotesize
\begin{tabular}{p{2.5cm}p{5cm}p{2.5cm}p{2.5cm}}
  \hline
 & WoS field & \# citations (techn) & \# citations (total) \\ 
  \hline
Solar PV & Eng., Electrical \& Electronic & 20,906 & 24,645 \\ 
  Wind & Eng., Electrical \& Electronic & 1,886 & 24,645 \\ 
  Solar thermal & Energy \& Fuels & 1,928 & 33,343 \\ 
  Energy from sea & Energy \& Fuels & 246 & 33,343 \\ 
  Hydroelectric & Energy \& Fuels & 154 & 33,343 \\ 
  Geothermal & Energy \& Fuels & 253 & 33,343 \\ 
  Biofuels & Biochemistry \& Molecular Biology & 70,729 & 89,764 \\ 
  Fuels from waste & Biochemistry \& Molecular Biology & 16,372 & 89,764 \\ 
  Nuclear fission & Nuclear Science \& Techn. & 1,228 & 1,511 \\ 
  Nuclear fusion & Physics, Fluids \& Plasmas & 1,360 & 2,192 \\ 
  \hline
  All & Biochemistry \& Molecular Biology & 89,764 & 89,764 \\ 
   \hline
\end{tabular}
\caption{Most cited WoS fields. Column (2) and (3) show the number of citations received within the technology type and from all types.} 
\label{tab:fiels_most_cited}
\end{table}

\begin{table}[ht]
\centering \footnotesize
\begin{tabular}{p{2.25cm}p{7.5cm}p{0.5cm}p{2cm}p{2cm}}
  \hline
 & Name &  & \# cit. (techn) & \# cit. (total) \\ 
  \hline
Solar PV & Applied Physics Letters & (J) & 11,423 & 11,624 \\ 
  Wind & Power Electronics Specialists Conference & (C) & 235 & 1,414 \\ 
  Solar thermal & Solar Energy & (J) & 490 & 1,303 \\ 
  Energy from sea & OCEANS Conference & (C) &  34 &  48 \\ 
  Hydroelectric & Journal of Fluid Mechanics & (J) &  16 &  87 \\ 
  Geothermal & Water Science and Technology & (J) &  44 & 513 \\ 
  Biofuels & Proceedings of the National Academy of Sciences of the USA & (J) & 8,100 & 10,723 \\ 
  Fuels from waste & Journal of Bacteriology & (J) & 2,385 & 9,124 \\ 
  Nuclear fission & Journal of Nuclear Materials & (J) & 179 & 234 \\ 
  Nuclear fusion & Nuclear Fusion & (J) & 361 & 388 \\ 
  \hline
  All & Applied Physics Letters & (J) & 11,624 & 11,624 \\ 
  \hline
\end{tabular}
\caption{Most cited journal (J) or conference proceeding (C) and number of citations received within technology type and from all types.} 
\label{tab:journal_most_cited}
\end{table}

\newpage

{\footnotesize
\begin{longtable}{|p{.2cm}|p{1.45cm}|p{6cm}|p{.65cm}|p{.65cm}|p{.65cm}|p{.75cm}|p{.75cm}|p{.75cm}|}
  \hline
& CPC class & Title & First & Last & Pat. & Sci. cites & Pat. cites & FW cites \\ 
  \hline
1 & Y02E10/00 & Energy generation through renewable energy sources & 2014 & 2018 &    32 &     65 &     18 &      7 \\ 
  2 & Y02E10/10 & Geothermal energy & 1900 & 2019 &   292 &    144 &  2,645 &  3,249 \\ 
  3 & Y02E10/12 & Earth coil heat exchangers & 1931 & 2019 &   154 &    257 &  2,932 &  1,883 \\ 
  4 & Y02E10/125 & Compact tube assemblies & 1951 & 2019 &   155 &     52 &  3,154 &  2,115 \\ 
  5 & Y02E10/14 & Systems injecting medium directly into ground & 1938 & 2018 &   163 &    286 &  2,390 &  2,540 \\ 
  6 & Y02E10/16 & Systems injecting medium into a closed well & 2015 & 2018 &     3 &      2 &     49 &      2 \\ 
  7 & Y02E10/20 & Hydro energy & 1841 & 2019 &   320 &     31 &  1,643 &  1,699 \\ 
  8 & Y02E10/22 & Conventional & 1859 & 2019 &   402 &     28 &  3,042 &  2,999 \\ 
  9 & Y02E10/223 & Turbines or waterwheels & 1837 & 2019 & 1,581 &     18 &  3,186 &  4,026 \\ 
  10 & Y02E10/226 & Other parts or details & 1836 & 2018 &   860 &     31 &  2,157 &  2,905 \\ 
  11 & Y02E10/28 & Tidal stream or damless hydropower & 1840 & 2019 & 1,444 &    542 & 15,576 &  8,203 \\ 
  12 & Y02E10/32 & Oscillating water column [OWC] & 1985 & 2018 &     5 &      0 &     38 &     21 \\ 
  13 & Y02E10/34 & Ocean thermal energy conversion [OTEC] & 1933 & 2019 &   239 &    203 &  3,675 &  2,520 \\ 
  14 & Y02E10/36 & Salinity gradient & 1985 & 2018 &    22 &    160 &    334 &    151 \\ 
  15 & Y02E10/38 & Wave energy or tidal swell & 1842 & 2019 & 1,709 &    872 & 20,492 & 15,835 \\ 
  16 & Y02E10/40 & Solar thermal energy & 1883 & 2019 & 1,548 &  1,535 & 15,115 & 23,024 \\ 
  17 & Y02E10/41 & Tower concentrators & 1904 & 2019 &   257 &    588 &  4,206 &  3,205 \\ 
  18 & Y02E10/42 & Dish collectors & 1955 & 2017 &    93 &    116 &  1,167 &  1,045 \\ 
  19 & Y02E10/44 & Heat exchange systems & 1881 & 2019 & 2,219 &    765 & 21,626 & 28,221 \\ 
  20 & Y02E10/45 & Trough concentrators & 1931 & 2017 &    84 &     31 &    614 &  1,041 \\ 
  21 & Y02E10/46 & Conversion of thermal power into mechanical power & 1880 & 2019 &   643 &    625 & 11,213 &  5,262 \\ 
  22 & Y02E10/465 & Thermal updraft & 1926 & 2019 &   138 &     54 &  1,522 &  1,657 \\ 
  23 & Y02E10/47 & Mountings or tracking & 1861 & 2019 & 2,007 &    765 & 39,241 & 25,031 \\ 
  24 & Y02E10/50 & Photovoltaic [PV] energy & 1926 & 2019 & 6,258 & 20,620 & 77,714 & 67,616 \\ 
  25 & Y02E10/52 & PV systems with concentrators & 1941 & 2019 & 1,720 &  6,054 & 25,708 & 17,531 \\ 
  26 & Y02E10/541 & CuInSe2 material PV cells & 1976 & 2019 &   863 &  5,992 & 13,880 &  9,309 \\ 
  27 & Y02E10/542 & Dye sensitized solar cells & 1974 & 2019 &   707 &  5,203 &  4,750 &  5,849 \\ 
  28 & Y02E10/543 & Solar cells from Group II-VI materials & 1966 & 2019 &   298 &  1,431 &  4,776 &  3,704 \\ 
  29 & Y02E10/544 & Solar cells from Group III-V materials & 1959 & 2019 & 1,098 &  9,652 & 15,669 & 12,934 \\ 
  30 & Y02E10/545 & Microcrystalline silicon PV cells & 1998 & 2015 &    36 &    101 &    532 &    284 \\ 
  31 & Y02E10/546 & Polycrystalline silicon PV cells & 1967 & 2019 &   192 &    502 &  1,473 &  1,830 \\ 
  32 & Y02E10/547 & Monocrystalline silicon PV cells & 1957 & 2019 & 1,779 & 16,106 & 21,580 & 19,892 \\ 
  33 & Y02E10/548 & Amorphous silicon PV cells & 1961 & 2019 &   510 &  1,296 &  3,512 &  9,951 \\ 
  34 & Y02E10/549 & organic PV cells & 1960 & 2019 & 3,566 & 34,932 & 29,881 & 17,013 \\ 
  35 & Y02E10/56 & Power conversion electric or electronic aspects & 1967 & 2019 &   157 &    926 &  5,619 &  1,802 \\ 
  36 & Y02E10/563 & for grid-connected applications & 1984 & 2019 &   838 &  3,122 & 11,553 &  7,261 \\ 
  37 & Y02E10/566 & concerning power management inside the plant & 1973 & 2019 &   187 &     99 &  1,674 &  3,369 \\ 
  38 & Y02E10/58 & Maximum power point tracking [MPPT] systems & 1965 & 2019 &   350 &  2,025 &  7,539 &  6,405 \\ 
  39 & Y02E10/60 & Thermal-PV hybrids & 1960 & 2018 &   127 &    371 &  1,573 &  1,299 \\ 
  40 & Y02E10/70 & Wind energy & 1853 & 2018 &   232 &     52 &  1,044 &  1,360 \\ 
  41 & Y02E10/72 & Wind turbines with rotation axis in wind direction & 1840 & 2019 & 1,658 &    307 &  7,820 &  6,068 \\ 
  42 & Y02E10/721 & Blades or rotors & 1859 & 2019 & 1,187 &    541 & 10,341 &  6,717 \\ 
  43 & Y02E10/722 & Components or gearbox & 1850 & 2019 &   318 &    153 &  2,896 &  1,470 \\ 
  44 & Y02E10/723 & Control of turbines & 1837 & 2019 & 1,399 &    956 &  7,763 &  7,871 \\ 
  45 & Y02E10/725 & Generator or configuration & 1882 & 2019 &   900 &    293 & 14,721 &  7,403 \\ 
  46 & Y02E10/726 & Nacelles & 1982 & 2019 &   119 &     36 &    836 &    396 \\ 
  47 & Y02E10/727 & Offshore towers & 1883 & 2019 &   133 &     31 &  1,517 &    504 \\ 
  48 & Y02E10/728 & Onshore towers & 1842 & 2019 &   590 &    133 &  7,145 &  4,325 \\ 
  49 & Y02E10/74 & Wind turbines with rotation axis perpend. to wind direction & 1840 & 2019 & 1,356 &    155 &  9,955 & 10,184 \\ 
  50 & Y02E10/76 & Power conversion electric or electronic aspects & 2001 & 2019 &    35 &    111 &    477 &    332 \\ 
  51 & Y02E10/763 & for grid-connected applications & 1985 & 2019 &   539 &  1,332 &  7,663 &  4,551 \\ 
  52 & Y02E10/766 & concerning power management inside the plant & 1979 & 2019 &    51 &     14 &    331 &     55 \\ 
  53 & Y02E30/00 & Energy generation of nuclear origin & 2017 & 2017 &     4 &      6 &      6 &      0 \\ 
  54 & Y02E30/10 & Fusion reactors & 1962 & 2018 &    32 &    155 &    282 &    328 \\ 
  55 & Y02E30/122 & Tokamaks & 1961 & 2018 &    55 &    412 &    871 &    532 \\ 
  56 & Y02E30/124 & Stellarators & 1959 & 1987 &    11 &     20 &     14 &    110 \\ 
  57 & Y02E30/126 & Other reactors with MPC & 1949 & 2019 &    91 &    173 &    412 &    909 \\ 
  58 & Y02E30/128 & First wall, divertor, blanket & 1973 & 2018 &    69 &    167 &    515 &    416 \\ 
  59 & Y02E30/14 & Inertial plasma confinement & 1968 & 2019 &    25 &    144 &    224 &    366 \\ 
  60 & Y02E30/16 & Injection systems and targets & 1964 & 2019 &   119 &    474 &    880 &  1,657 \\ 
  61 & Y02E30/18 & Low temperature fusion & 1967 & 2018 &    52 &    505 &    609 &    490 \\ 
  62 & Y02E30/30 & Nuclear fission reactors & 1964 & 2014 &    15 &      9 &     43 &     98 \\ 
  63 & Y02E30/31 & Boiling water reactors & 1960 & 2018 &   160 &     49 &  1,074 &  1,006 \\ 
  64 & Y02E30/32 & Pressurized water reactors & 1961 & 2018 &   114 &     46 &  1,048 &    652 \\ 
  65 & Y02E30/33 & Gas cooled reactors & 1955 & 2018 &   116 &     40 &    434 &    772 \\ 
  66 & Y02E30/34 & Fast breeder reactors & 1909 & 2018 &   147 &    160 &  1,428 &    814 \\ 
  67 & Y02E30/35 & Liquid metal reactors & 1967 & 2017 &    47 &     21 &    317 &    426 \\ 
  68 & Y02E30/36 & Pebble bed reactors & 1957 & 2014 &    72 &     34 &    442 &    302 \\ 
  69 & Y02E30/37 & Accelerator driven reactors & 1966 & 1999 &     5 &      4 &     18 &     83 \\ 
  70 & Y02E30/38 & Fuel & 1909 & 2019 & 1,176 &    615 &  6,833 &  7,009 \\ 
  71 & Y02E30/39 & Control of nuclear reactions & 1887 & 2019 & 1,301 &    528 &  7,579 &  9,153 \\ 
  72 & Y02E30/40 & Other aspects relating to nuclear fission & 1838 & 2019 & 5,431 &  1,881 & 36,007 & 37,457 \\ 
  73 & Y02E50/00 & Technologies for the production of fuel of non-fossil origin & 2015 & 2015 &     2 &     12 &    102 &      3 \\ 
  74 & Y02E50/10 & Biofuels & 1855 & 2019 &   910 & 28,041 & 14,777 &  7,127 \\ 
  75 & Y02E50/11 & CHP turbines for biofeed & 1976 & 2018 &    36 &     48 &    543 &    342 \\ 
  76 & Y02E50/12 & Gas turbines for biofeed & 1973 & 2018 &   142 &    502 &  4,996 &  2,386 \\ 
  77 & Y02E50/13 & Bio-diesel & 1980 & 2019 &   968 & 30,718 & 16,088 &  6,918 \\ 
  78 & Y02E50/14 & Bio-pyrolysis & 1889 & 2019 &   588 &  3,378 &  8,193 &  4,462 \\ 
  79 & Y02E50/15 & Torrefaction of biomass & 1940 & 2018 &    16 &     13 &    156 &     54 \\ 
  80 & Y02E50/16 & Cellulosic bio-ethanol & 1927 & 2019 &   632 & 34,564 &  8,863 &  3,818 \\ 
  81 & Y02E50/17 & Grain bio-ethanol & 1869 & 2019 & 1,164 & 38,661 & 10,754 &  7,631 \\ 
  82 & Y02E50/18 & Bio-alcohols produced by other means than fermentation & 1981 & 2018 &   126 &    982 &  3,316 &    888 \\ 
  83 & Y02E50/30 & Fuel from waste & 1848 & 2019 & 1,188 &  4,778 & 17,322 &  9,238 \\ 
  84 & Y02E50/32 & Synthesis of alcohols or diesel from waste
  & 1975 & 2019 &   372 &  2,974 & 10,136 &  3,397 \\ 
  85 & Y02E50/343 & production by fermentation of organic by-products & 1919 & 2019 & 1,584 & 40,676 & 23,745 & 15,788 \\ 
  86 & Y02E50/346 & from landfill gas & 1979 & 2019 &   108 &    514 &  2,975 &  1,636 \\ 
   \hline 
\caption{Counts and citations statistics for LCETs at the highest possible resolution. Columns \textit{First} and \textit{Last} refer to the year of the earliest and most recent patents, respectively. \textit{Pat.} denotes the number of patents. \textit{Sci. cites} and \textit{Pat. cites} are the total number of scientific and patent citations made, respectively. \textit{FW cites} is the number of forward citations, i.e. the total of citations received by patents in a given class. 
Note that there was a major revision of Y02E classifications in August 2020, resulting in a less detailed breakdown as shown here. 
\label{t:greentech_details} }
\end{longtable}
}

\FloatBarrier

\section{Cleaning data on early patents}
\label{sec:si_cleaning_early_pats}

Our main result is that there is a rising importance of science in LCET patenting. To be sure that this result is not driven by artefacts in the data, we proceed to a significant manual checking and cleaning of the data.

Recent patents are written in digitized form natively. This makes them easy to parse. They also have a list of references on the front page, with references to patents in a subsection and references to other documents in a separate subsection. In contrast, scientific citations in earlier patents are much harder to parse, because the text itself comes from optical character recognition of old documents and is harder to decipher the further in time we go back, and because bibliographic references are within the body of the text. Fortunately, the RoS files include references to science that are found in the body of the text, but of course, these are likely to contain more mistakes.

At this point, it is important to note that 1976 marks a important change in the quality of the data. Before this date, most citations to science are found in the body of the text, and the RoS confidence scores are generally low. After this date, most citations to science are found in the reference list, and the RoS confidence scores are high.

This implies that data cleaning efforts should concentrate on early years, say up to 1976. For Solar thermal, Geothermal, Ocean and Wind, each patent was checked until (and excluding) 1976. For Hydro, this check was done until 1987, when a reasonable share of non-false positive patents began. In Fusion, the first patent citing papers is from 1959 and is parsed correctly. This is not checked further. In Fission and Solar PV, patents clearly rely on science and tend to make citations in a clear scientific style that presumably makes parsing easier. Fission was checked until 1956 and Solar PV until 1958, because the number of patents is too high to check manually.

Cleaning up earlier patents more than later patents creates a potentially important issue, as it automatically decreases the observed reliance of science of earlier periods without affecting recent periods. Thus, the result that LCET technologies rely more and more on science could be the result of the data cleaning procedure. However, our results were similar on the uncleaned data.

A more serious issue is if the RoS data set, rather than the additional data cleaning, is biased towards identifying scientific references more easily in recent patents, which is plausible. This is addressed in two ways. First of all, while reading patents manually for our own data cleaning described above, specific attention is paid to ``false negatives'', that is, references that are in the patents but not in the RoS data. While some instance of this were found, they were rare. Second, a more rigorous approach was used and all early patents of a specific technology in a given year were read. Wind in 1875 chosen. All 52 patents were inspected, and it was confirmed that none of them had a citation to a scientific paper. Incidentally, it is striking to compare the 1875 patents with the 20th century ones. Early patents, indeed, do \emph{not} rely very much on academic knowledge, but instead describe the successful experiments of independent inventors at improving mechanical designs for windmills. In contrast, the 20th century patents do rely on academic knowledge, with clear references to knowledge in aerodynamics, electrical engineering, etc. 

\section{Innovation dynamics of Renewables, Nuclear power and Non-fossil fuels}
\label{SI:fig:timeseries_nuc_share}
\begin{figure}
	\centering
	\includegraphics[width=0.8\textwidth]{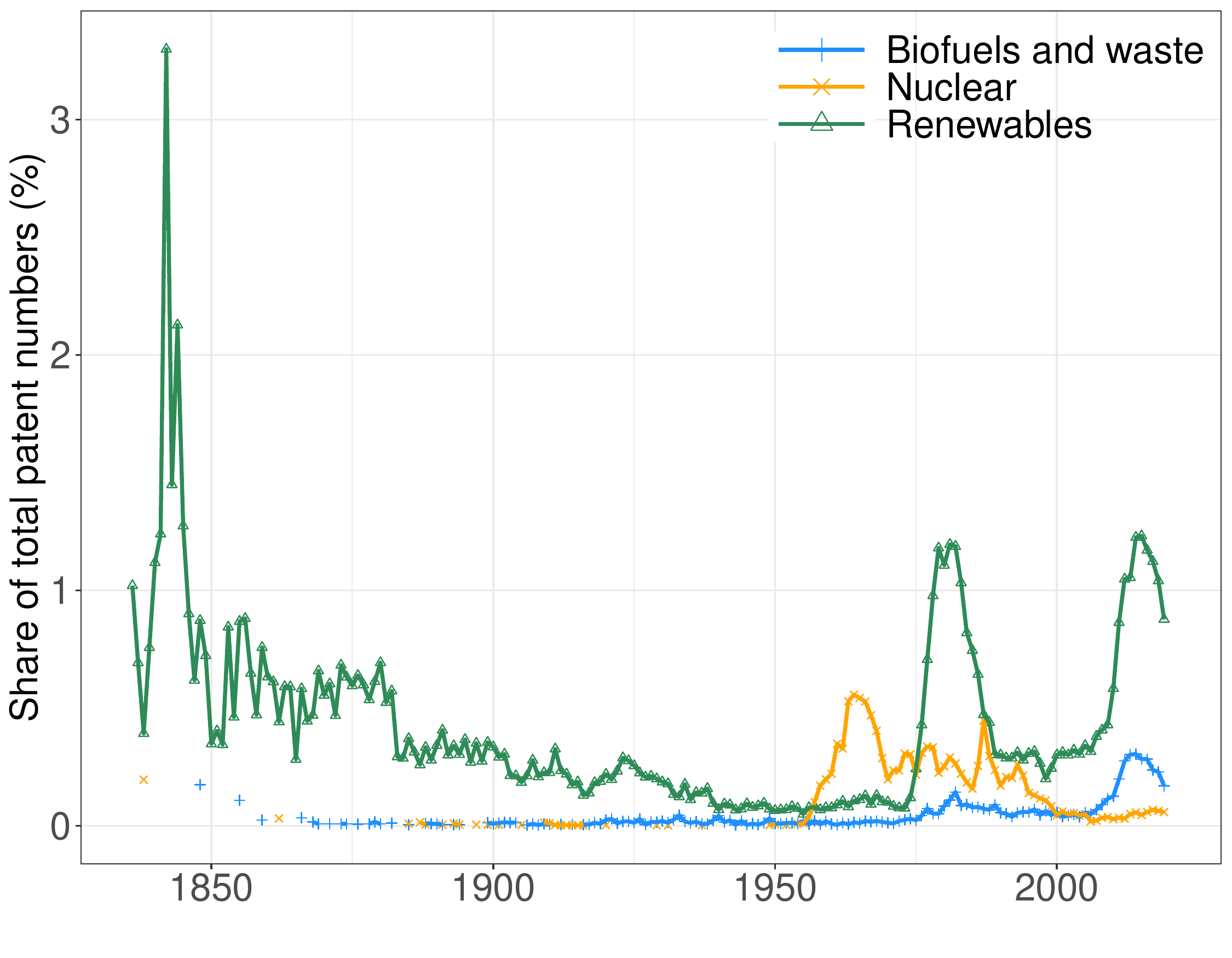}
	\caption{Share of LCET patents as a fraction of total number of patents over time, separating Renewables (Y02E10),
	Nuclear (Y02E30) and
	Biofuels and Waste (Y02E50).}
	\label{fig:timeseries_nuc_share}
\end{figure}

Fig. \ref{fig:timeseries_nuc_share} complements Fig. \ref{fig:full_timeseries} by showing the share of LCETs in total patenting based on three broader groups. 
The figure shows that patenting in low-carbon energy technology is old, but has fluctuated widely, particularly reacting to competition from fossil fuels \cite[cf.][]{unruh2000understanding}. In fact, the share of Renewables in total patenting was at a historical \emph{low} in the first half of the 20th century, picked up sharply after the oil crisis before falling back and picking up again starting in around 2005, to sit at 1\% now.

\section{Reliance on scientific fields} 
\label{sec:SI_networks}

\begin{figure}
\centering
\includegraphics[trim = {.8cm 0cm .8cm 0cm}, clip,width=.42\textwidth]{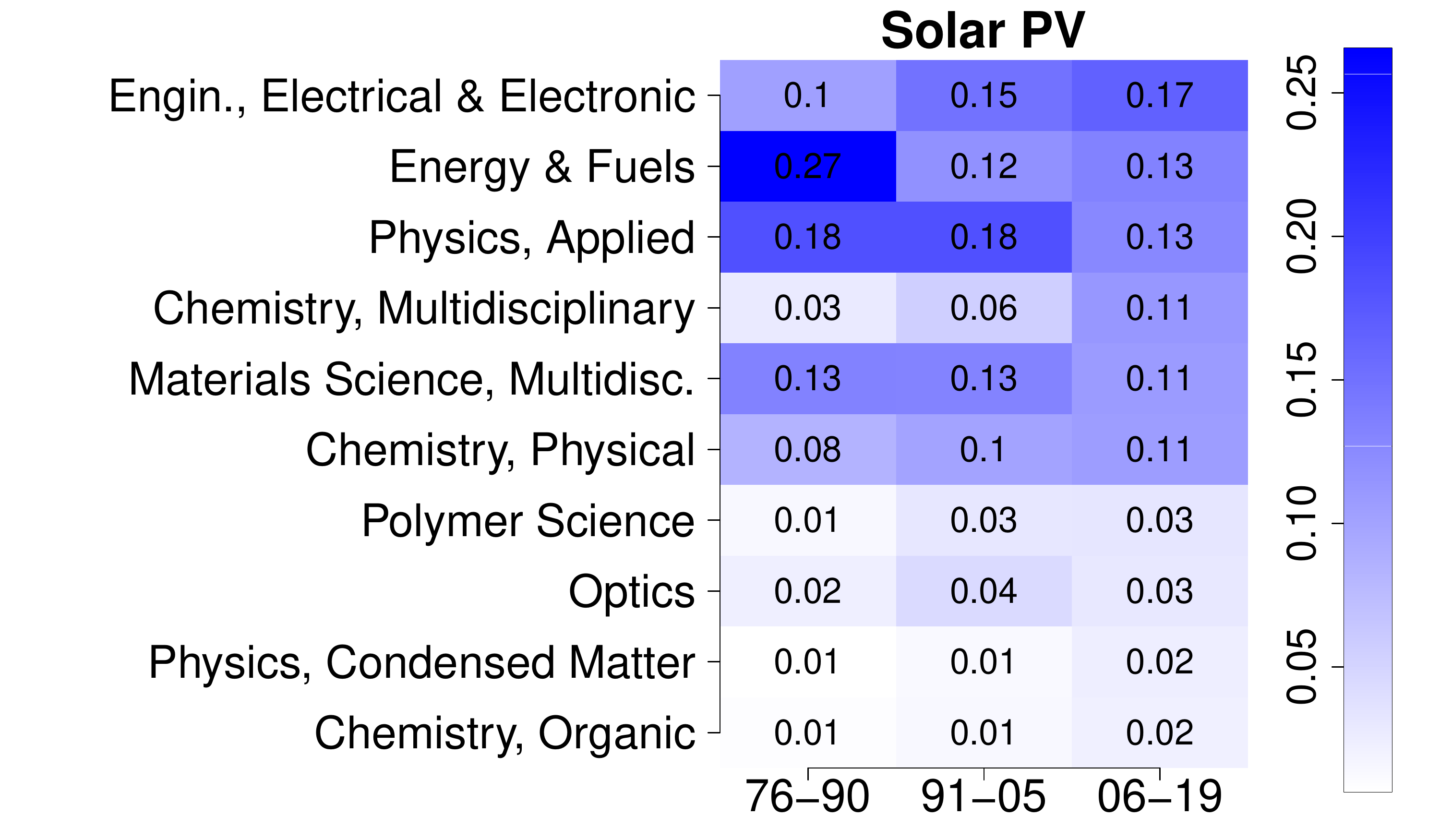}
\includegraphics[trim = {.8cm 0cm .8cm 0cm}, clip,width=.42\textwidth]{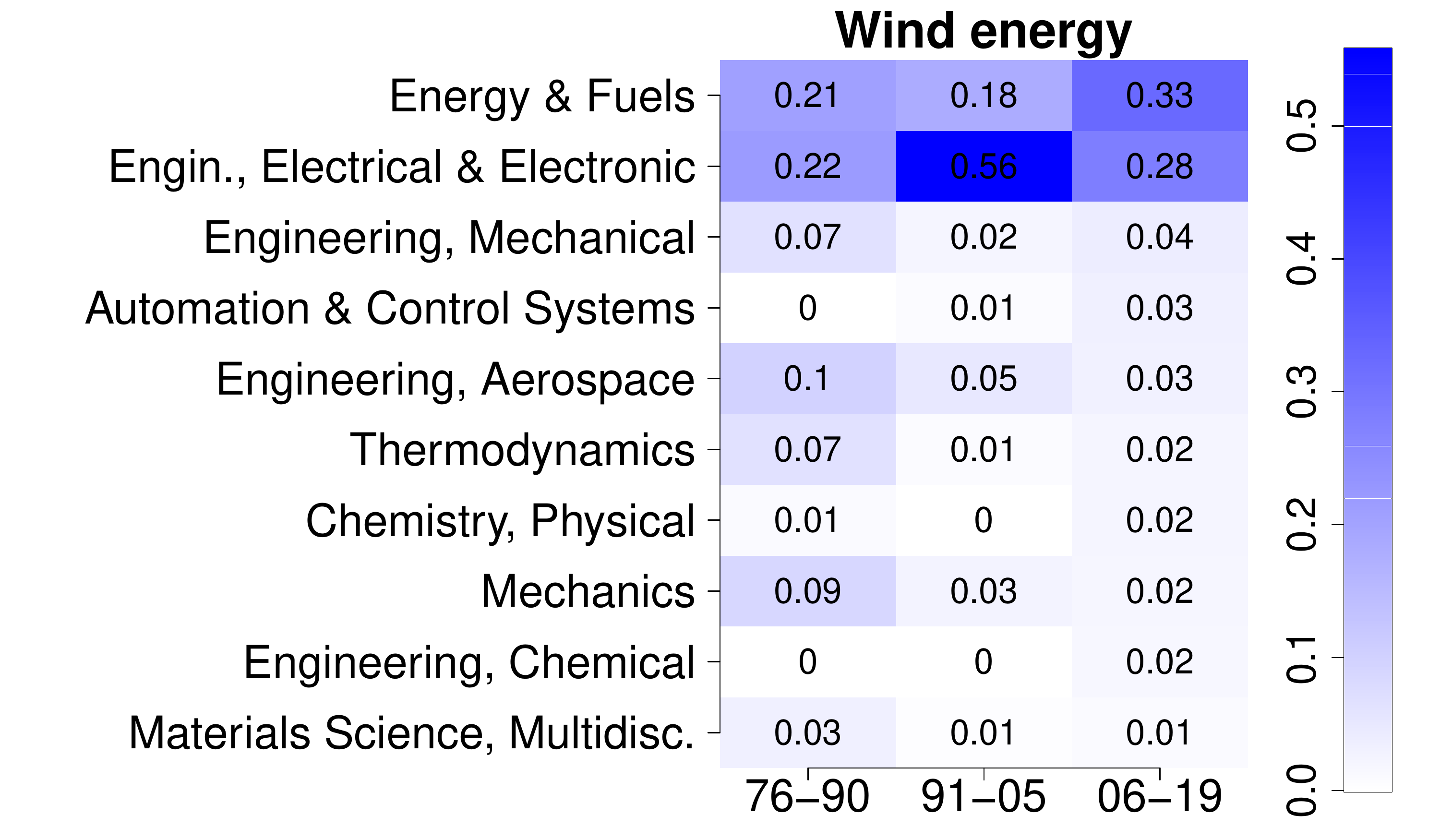}
\includegraphics[trim = {.8cm 0cm .8cm 0cm}, clip,width=.42\textwidth]{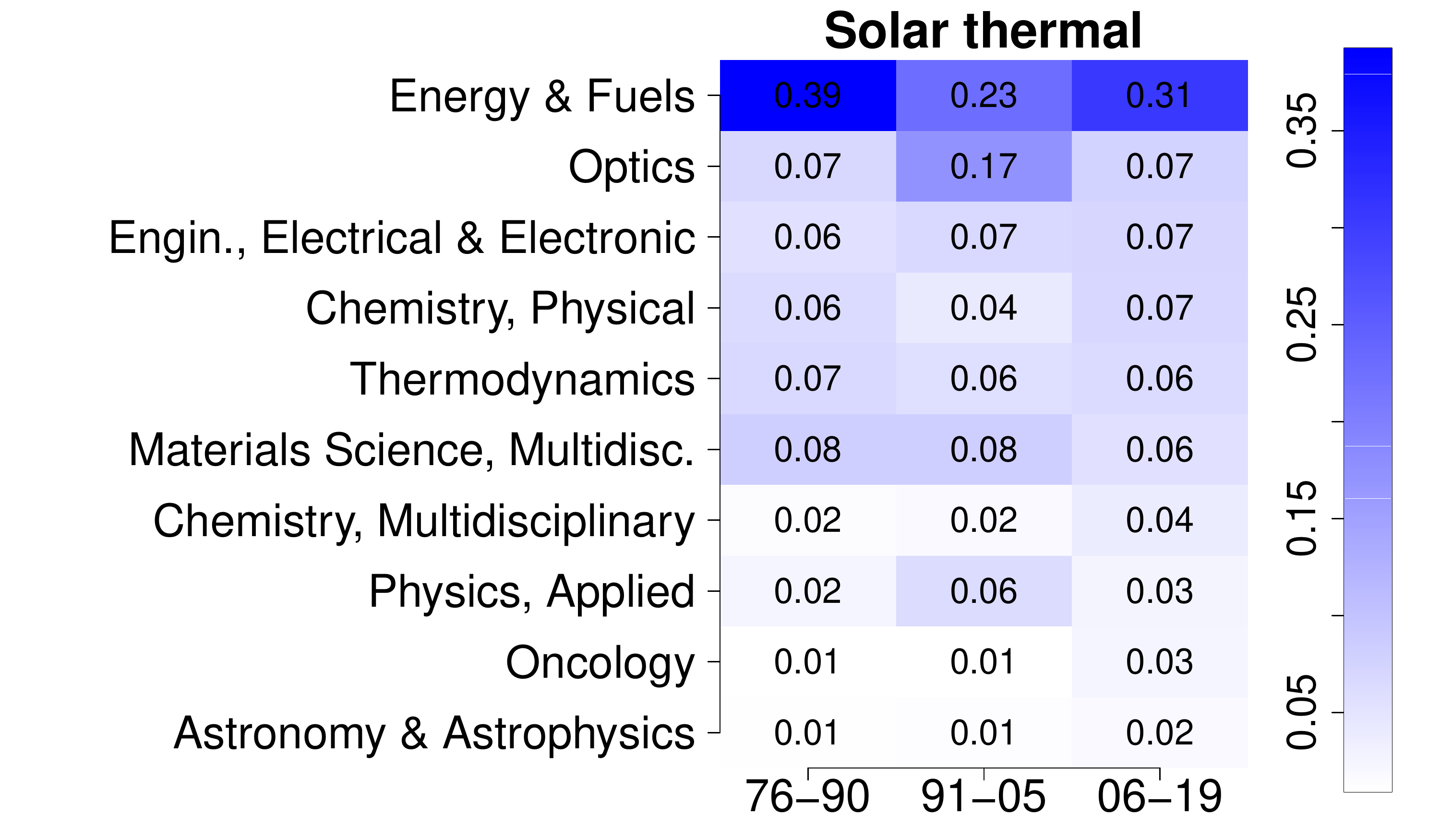}
\includegraphics[trim = {.8cm 0cm .8cm 0cm}, clip,width=.42\textwidth]{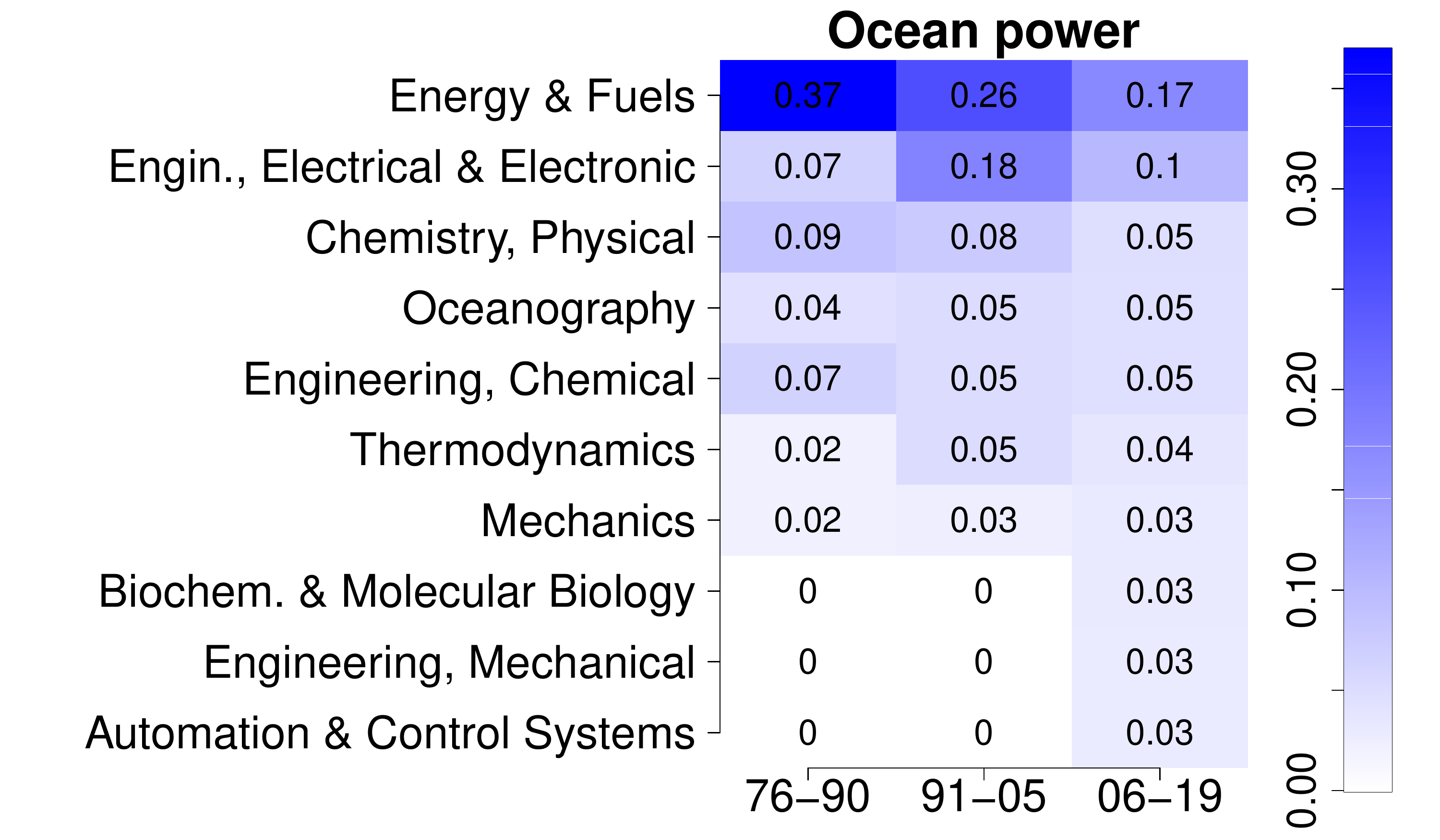}
\includegraphics[trim = {.8cm 0cm .8cm 0cm}, clip,width=.42\textwidth]{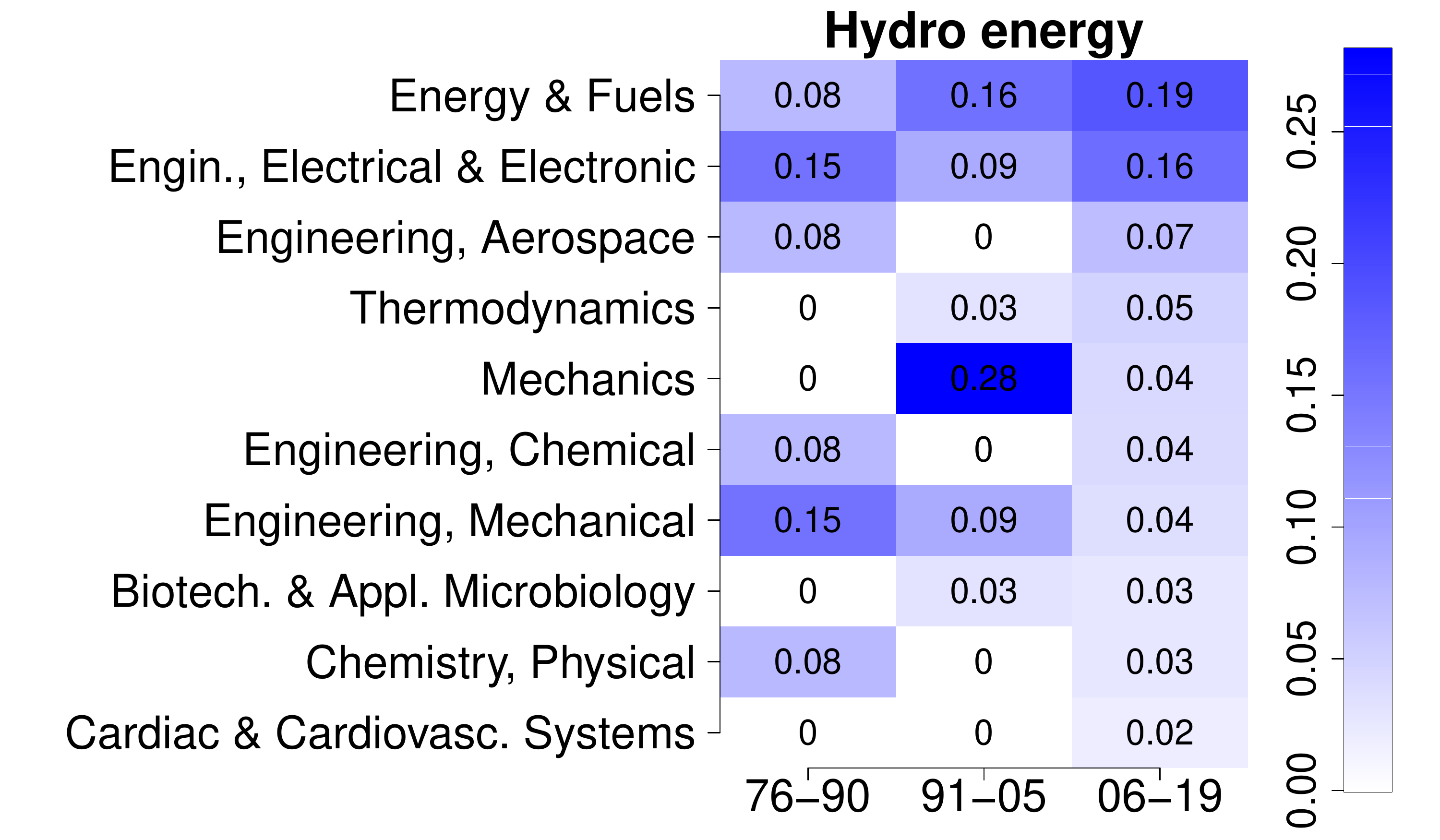}
\includegraphics[trim = {.8cm 0cm .8cm 0cm}, clip,width=.42\textwidth]{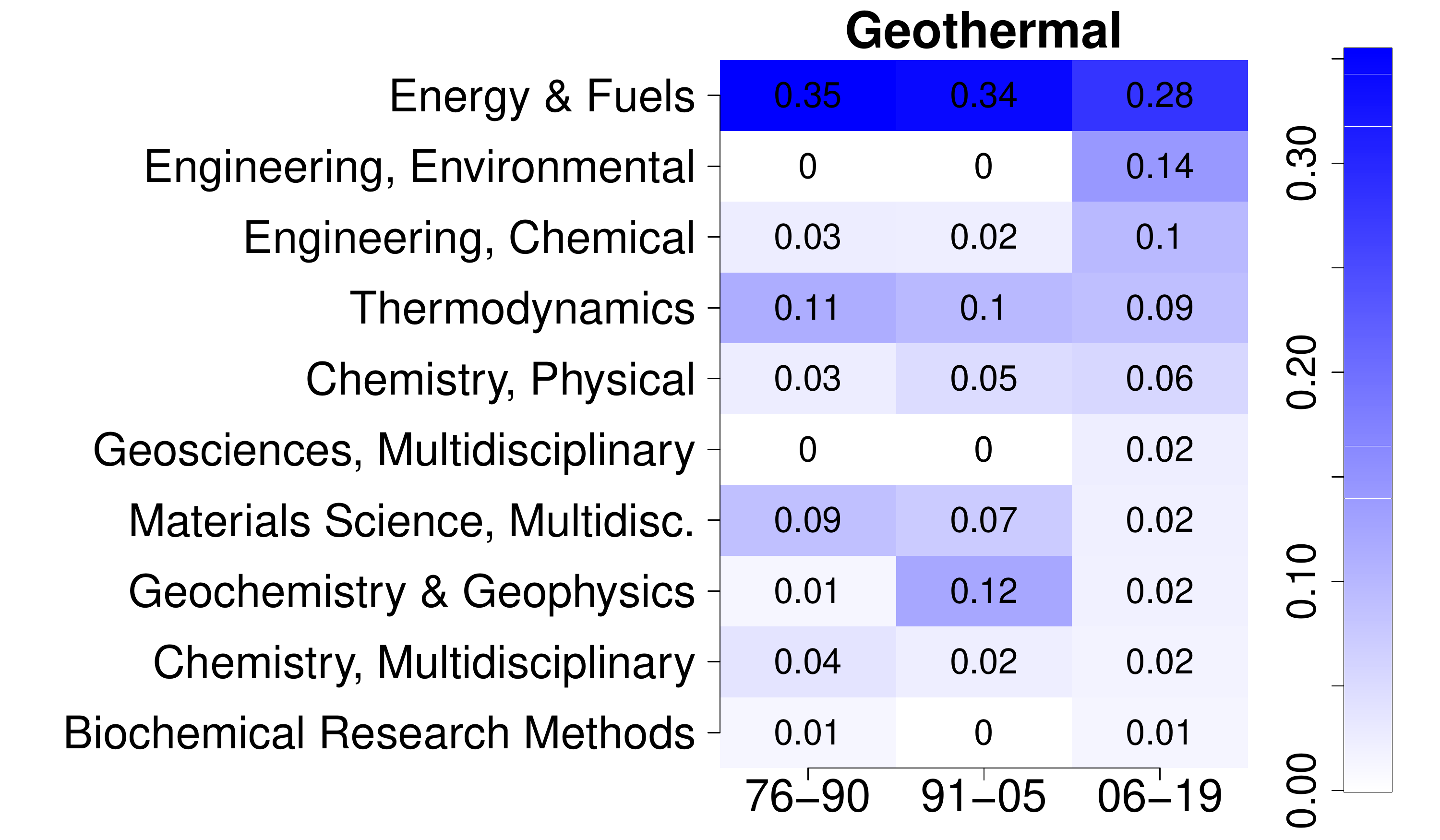}
\includegraphics[trim = {.8cm 0cm .8cm 0cm}, clip,width=.42\textwidth]{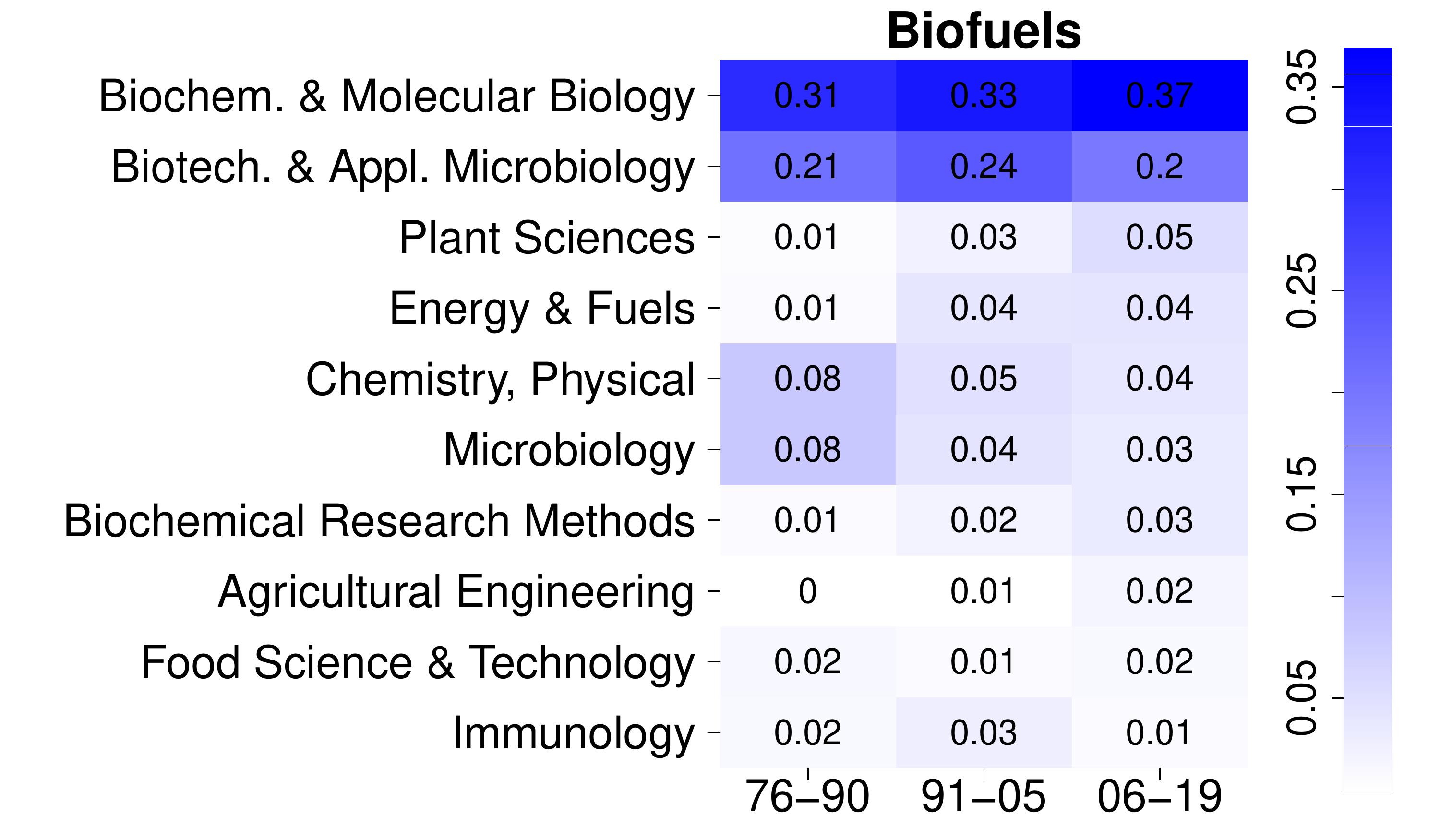}
\includegraphics[trim = {.8cm 0cm .8cm 0cm}, clip,width=.42\textwidth]{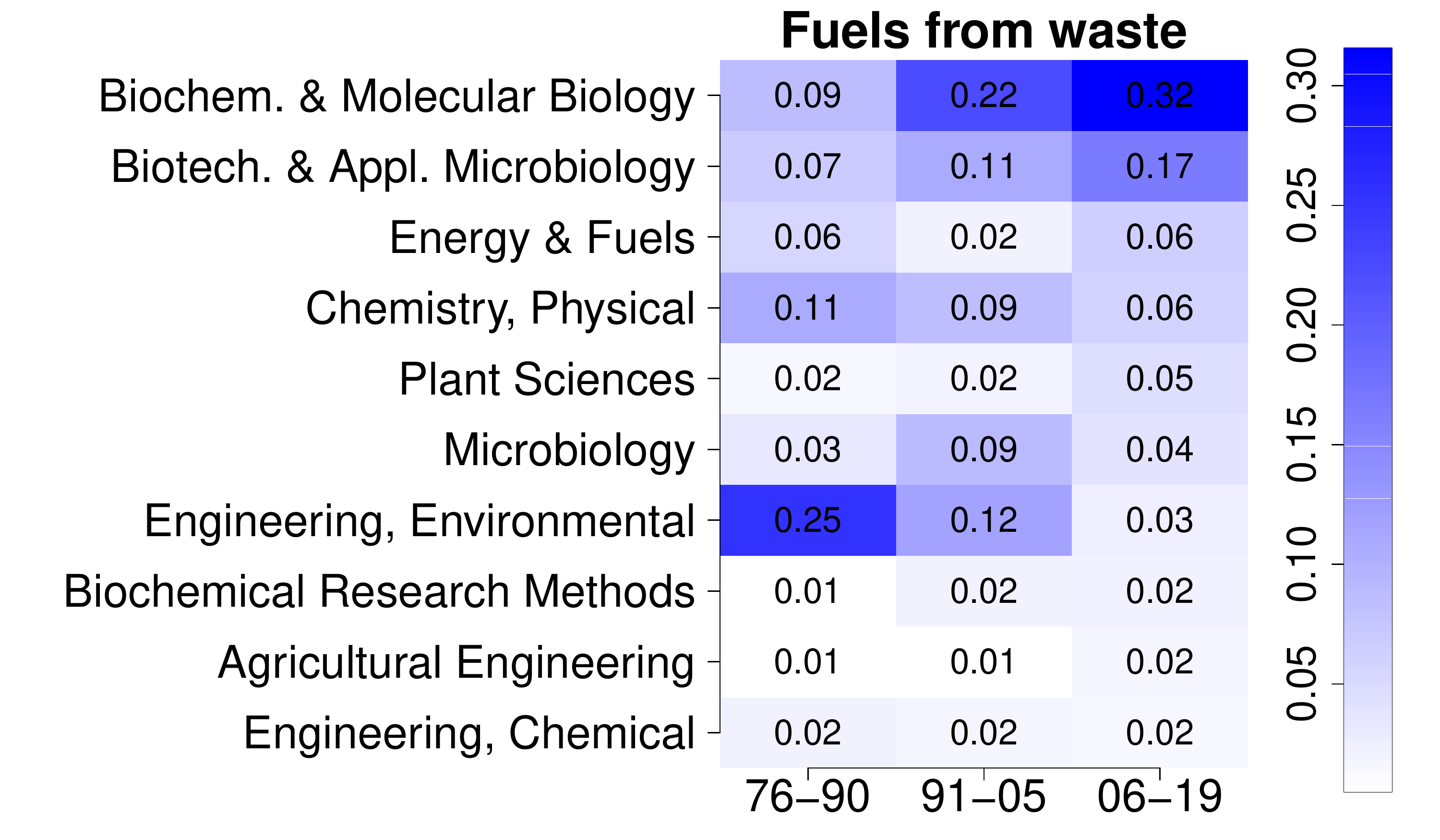}
\includegraphics[trim = {.8cm 0cm .8cm 0cm}, clip,width=.42\textwidth]{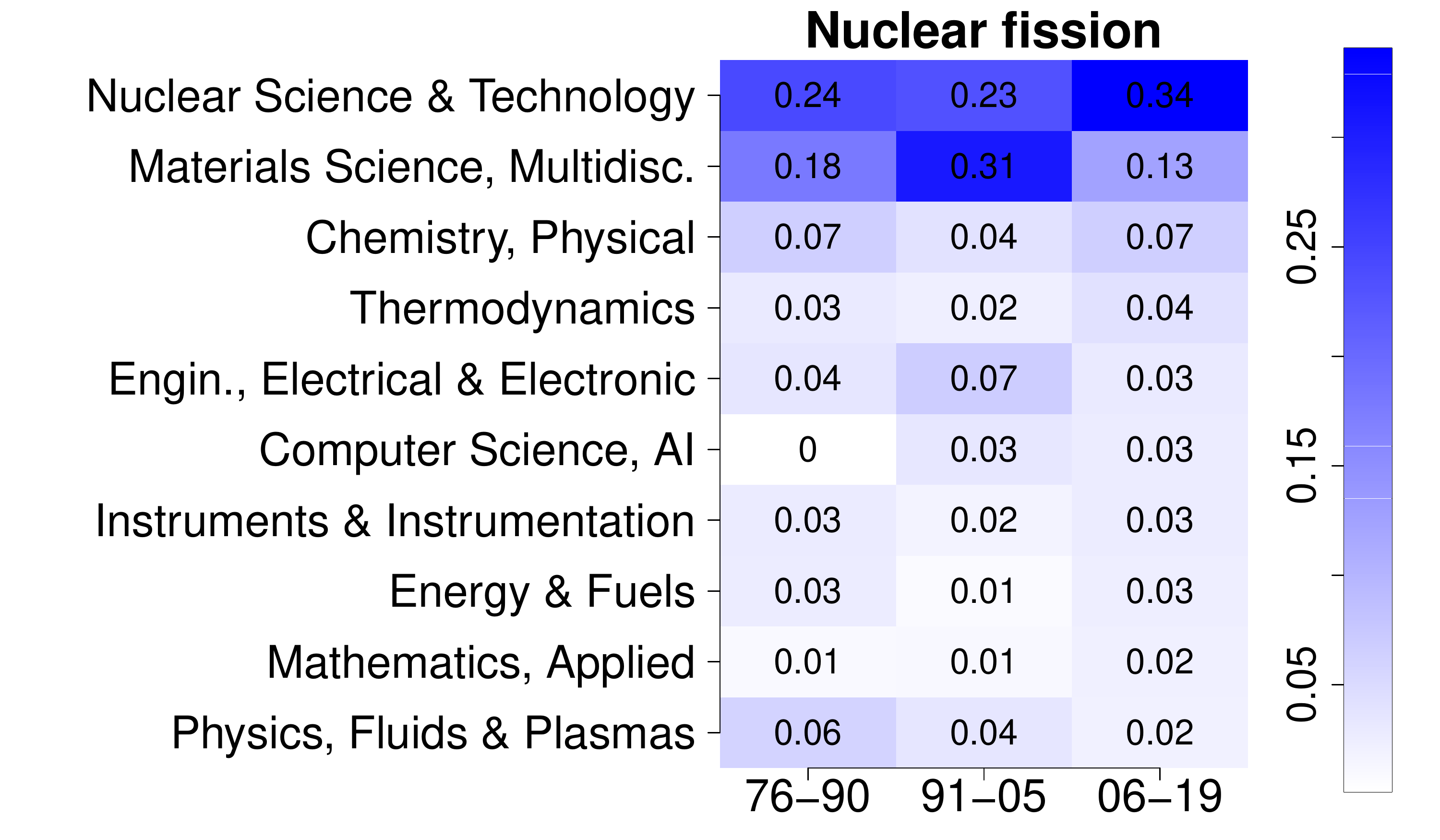}
\includegraphics[trim = {.8cm 0cm .8cm 0cm}, clip,width=.42\textwidth]{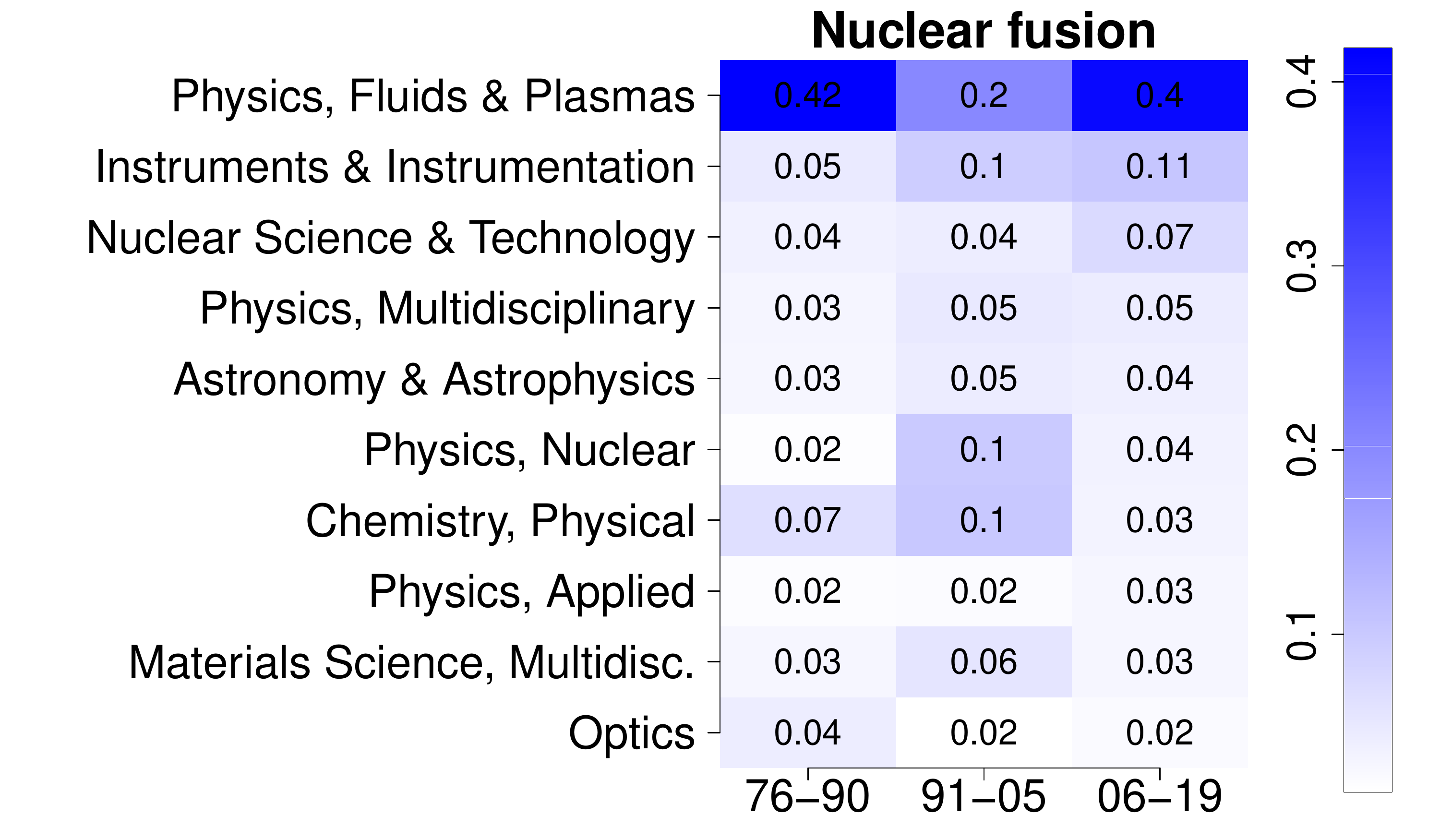}
    \caption{The ten most important scientific fields per LCET measured as shares of total references citing the field over time. Scientific fields are ordered from the most to least important field of the most recent time period.}
    \label{fig:SI_science_reliance_change}
\end{figure}

Fig. \ref{fig:SI_science_reliance_change} visualizes the most important WoS field per LCET category, as in Fig. \ref{fig:green_reliance_on_wos} of the main text but for the ten most important fields per LCET. WoS fields are ordered according to the most important field in the last time period. As can be seen from the main text, technologies are highly heterogeneous with respect to their scientific dependence. Some technologies experience pronounced changes in the citation structure over time, while others remain very stable. For example, Geothermal relies heavily on ``Energy \& Fuels'' across all time periods, but increasingly cites ``Environmental Engineering". Environmental Engineering becomes the second most important field in the last time period, although it didn't play any role before.

There is large time variations in Hydro energy but it should be recalled that there were only very few citations to scientific fields in the first two time horizons. In recent years the scientific reliance of Hydro is fairly evenly distributed. In contrast, technologies such as Wind, Fusion and Biofuels are heavily centred around only one or two scientific fields.
The citation distribution of Solar PV is remarkably flat. While citations from Solar PV were heavily concentrated on ``Energy \& Fuels'' in the early years, there is now a set of six scientific fields which the technologies relies on in very similar proportions. 

\begin{figure}
\centering
\includegraphics[trim = {0cm 0cm 3.9cm 0cm}, clip, width=.49\textwidth]{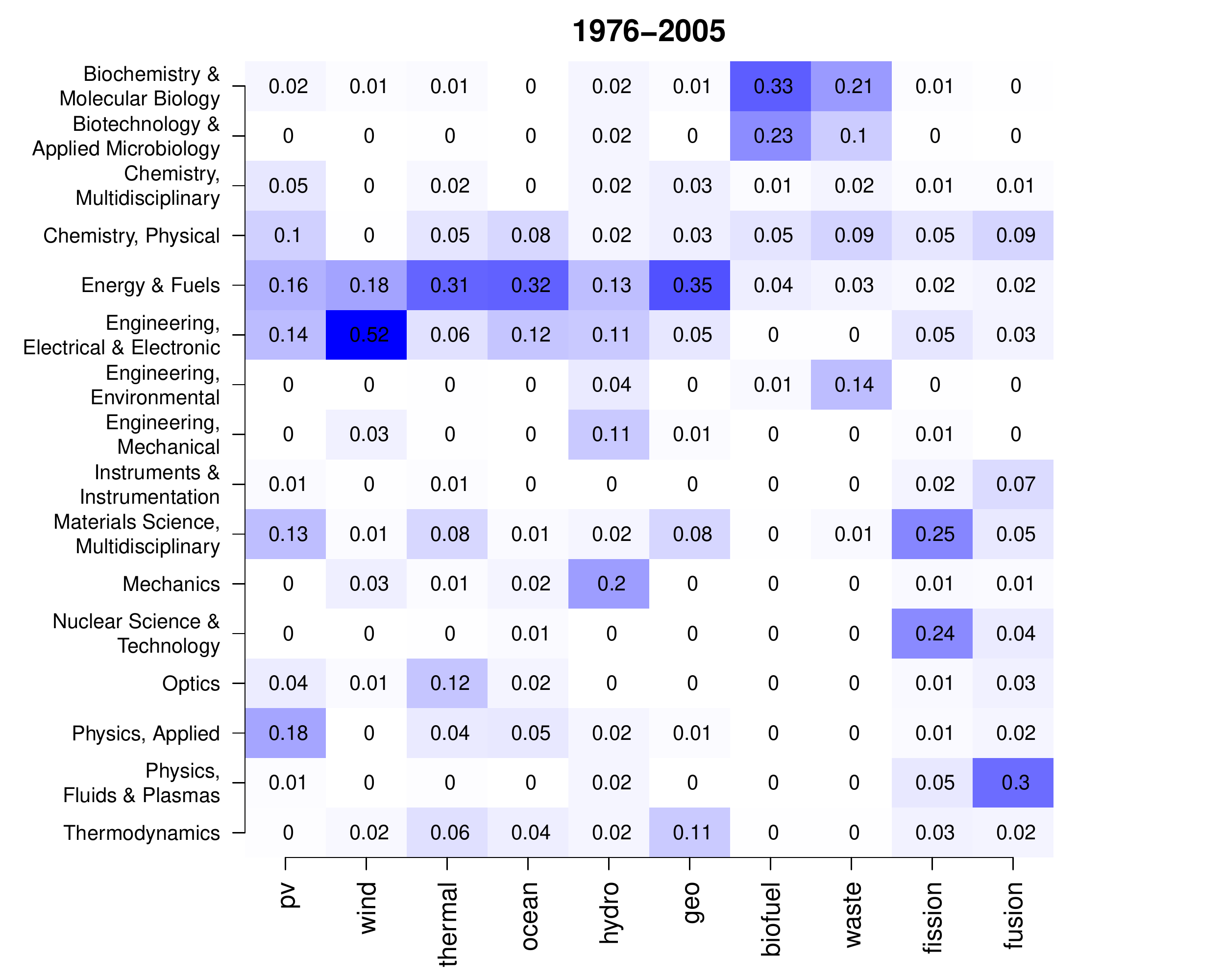}
\includegraphics[trim = {3.9cm 0cm 0cm 0cm}, clip, width=.49\textwidth]{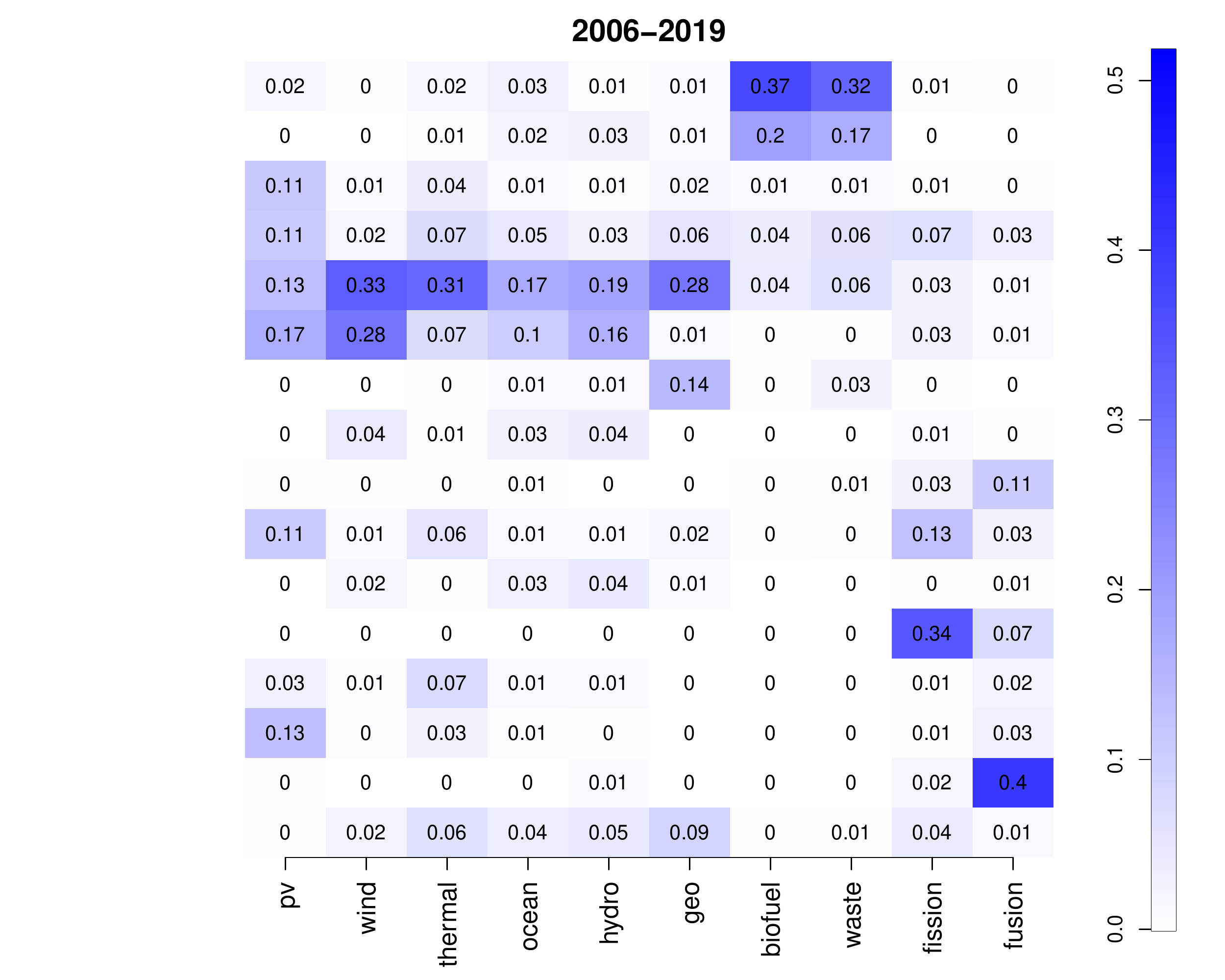}
\caption{Shares of technology-scientific field citations for all WoS fields which account for at least 10\% of a LCETs' overall citations.}
\label{fig:SI_science_similarity_matrix}
\end{figure}

Fig. \ref{fig:SI_science_similarity_matrix} shows the relative citations shares from LCETs to WoS fields for every field which accounts for at least 10\% of one technology's overall citations.
It becomes clear that non-fuel Renewable energy technologies rely fairly strongly on ``Energy \& Fuels'', in particular in the early time period, with ``Electrical Engineering and Electronics'' also playing an important role. 
``Energy \& Fuels'' and ``Electrical Engineering \& Electronics'' play a less important role for Nuclear power and non-fossil fuel technologies, which tend to build upon very different fields. Biofuels and Waste are focused on Biochemistry and Biotechnology, whereas Plasma physics is the most relevant field for Fusion patents and ``Nuclear Science \& Technology'' for Fission.

\FloatBarrier
\section{Normalized bibliographic coupling} \label{sec:SI_bibcoup}

\begin{figure}
	\centering
	\includegraphics[width = .8\textwidth, trim = {15cm .5cm 0.3cm .5cm}, clip]{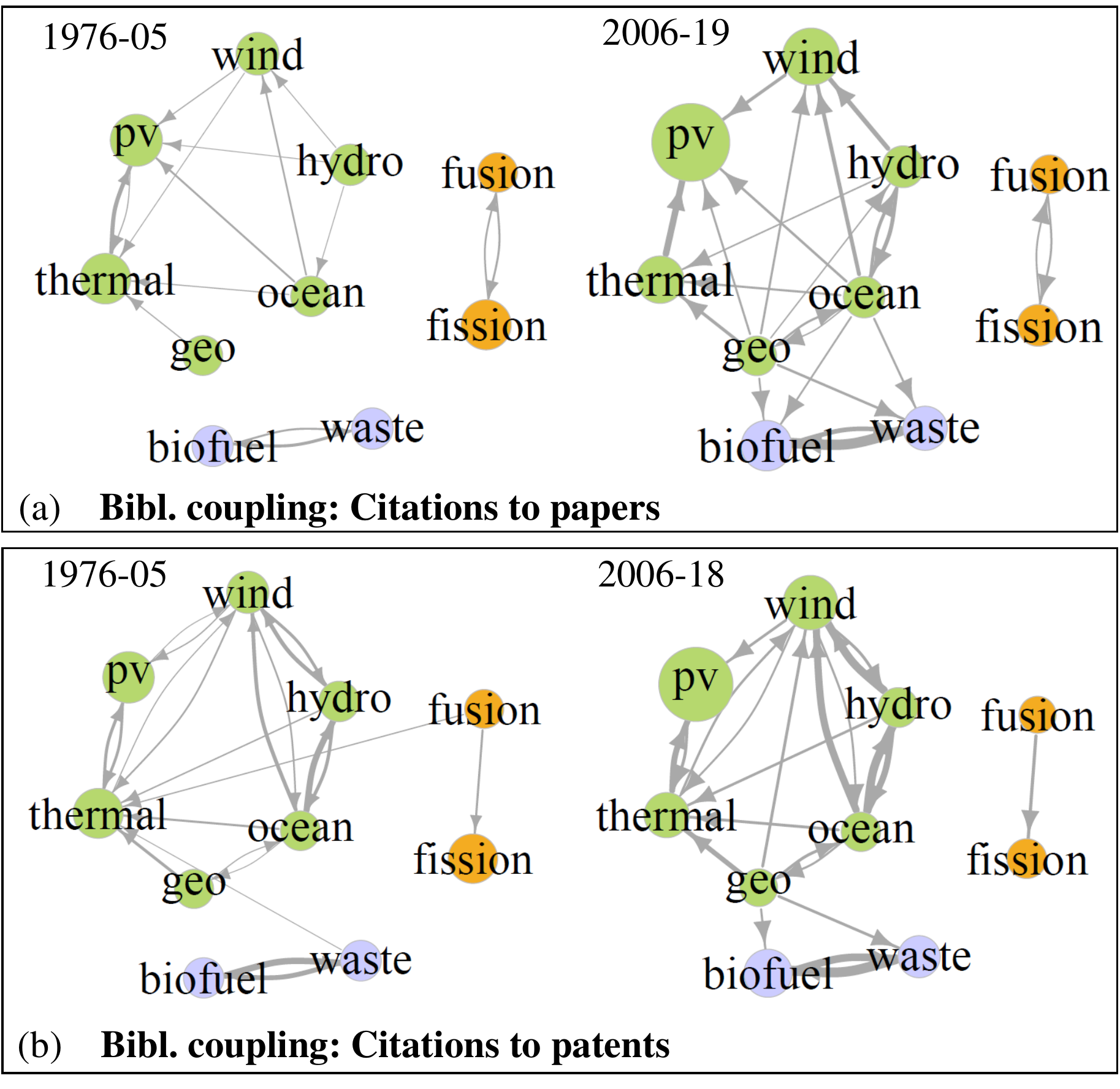}
	\caption{
	Normalized bibliographic coupling networks based on (a) citations to papers and (b) citations to patents. Bibliographic coupling measures how often a LCET cites the same paper and patent, respectively. Links are normalized by the total number of papers/patents cited and thus directed. A weighted directed link from $i$ to $j$ indicates the share of papers/patents cited by $i$ which are also cited by $j$. Only the largest quarter of links is shown.}
	\label{fig:SI_bibcoup}
\end{figure}

In the main text, we quantified the similarity of technologies' citations to scientific fields and technological classes. To check if the results hold for alternative metrics of technological dependence, a second, more fine-grained measure is considered here: paper- and patent-based bibliographic coupling networks.

Paper-based bibliographic coupling refers to counting for a given pair of technologies the number of papers which are cited by both technologies. Instead of asking whether two technologies rely on similar scientific fields, this measures how often two technologies refer to the very same academic paper.
Following the notation in the main text, it can easily be verified that the paper-based bibliographic coupling is given by the one-mode projection $\tilde{D} = C C^\top$ where the binary matrix $C$ is defined as $C_{kj}= 1$, if $[B^\top M]_{kj} >0$ and $0$ else.
The diagonal elements of $\tilde{D}$ give the total number of distinct papers cited by the corresponding technologies. By normalizing each row in $\tilde{D}$ by its diagonal element, we obtain an asymmetric matrix (directed network) $D = \text{diag}(\tilde{D})^{-1} \tilde{D}$ where the elements in the $k^{\text{th}}$ row indicate the share of papers cited by technology $k$ that are also cited by other technologies.

Fig. \ref{fig:SI_bibcoup}(a) shows the normalized bibliographic coupling network where a large edge \emph{from} node $k_1$ \emph{to} $k_2$ indicates that a large fraction of papers cited by $k_1$ are also cited by $k_2$. 
Compared to the cosine similarity networks of the main text (Fig. \ref{fig:similarity}), the separation of clusters is even more extreme. 
None of the inter-cluster links between 1976 and 2005 are above the 25\%-cutoff to be depicted visually. In the latter period there are links from Geo and Ocean to Biofuels and Waste. Bioefuels (Waste) patents cite around 7\% (9\%) and 8\% (12\%) of papers which are also cited by Ocean and Geo, respectively. 
There is almost no overlap between Nuclear paper citations and citations of other technologies.
The strongest links between Nuclear and other technologies are from Solar PV which cited 2.4\% and 1.6\% of Fission-cited and Fusion-cited papers, respectively. These numbers became substantially smaller for the more recent time period.
The links from Biofuels to Waste, Solar thermal to Solar PV, Waste to Biofuels and between the triangle Hydro-Wind-Ocean are the strongest. 
These relationships are not always symmetric. For example, over 40\% of all patents cited by Solar thermal has also been cited by Solar PV in the most recent period. Conversely, only 3\% of PV-cited papers are also cited by Solar thermal patents. But note that the total number of patents and citations of Solar PV is substantially larger than for Solar thermal. 

Hydro and Wind energy are again interesting examples exhibiting substantial changes over time. There was almost no overlap between the 40 papers cited by Hydro technologies before 2005 and the papers cited by all other technologies. Similarly, only few overlaps for the 405 papers cited by Wind can be found in the same period. In sharp contrast to earlier years, 36\% of all papers cited by Hydro patents have also been cited by Wind energy patents since 2006. Also, 31\% of all papers cited by Ocean patents have been cited by Hydro patents in the more recent period. The links between Geothermal and Solar thermal became stronger, as well as those between Ocean and Wind. Similarly to the cosine similarity networks in the main text, the links within the Renewables cluster and within the Fuels cluster got stronger in time, which is not true for the Nuclear equivalent.

In the main text, we complemented the network analysis of citations to scientific fields by an analysis of citations to patent technological classes.
Similarly, the degree of bibliographic coupling can also be computed for patent citations instead of paper citations. A link in a patent-based bibliographic network quantifies how often two technological classes cite the same patents.

The patent-based bibliographic coupling is given by $\tilde{T} = S S^\top$ where the binary matrix $S$ is defined as $S_{ki}=1$, if $ [B^\top H]_{ki} >0$ and $0$ else.
We again normalize every row in $\tilde{T}$ by its diagonal elements to obtain the normalized asymmetric bibliographic coupling matrix $T$ where elements of the $k^{\text{th}}$ row giving the shares of patents cited by technology $k$ cited by other technologies.

Fig. \ref{fig:SI_bibcoup}(b) visualizes the directed patent-based bibliographic coupling network which is again separated into three loosely connected components.
The strongest links can again be found between Biofuels and Waste and between Hydro, Wind and Ocean energy technologies, followed by the edge from Solar thermal to Solar PV. The average weight of links between Renewable energy technologies has substantially increased, as well as the links between Biofuels and Waste. Here, even the links between the Nuclear power technologies have slightly increased over time.

\section{Robustness checks} \label{sec:SI_robust}

\subsection{Scientific citation confidence scores}
\label{sec:SI_robust_CS}

\begin{figure}
\centering
\includegraphics[trim = {.75cm 1.75cm 0cm 2.8cm}, clip, width=1\textwidth]{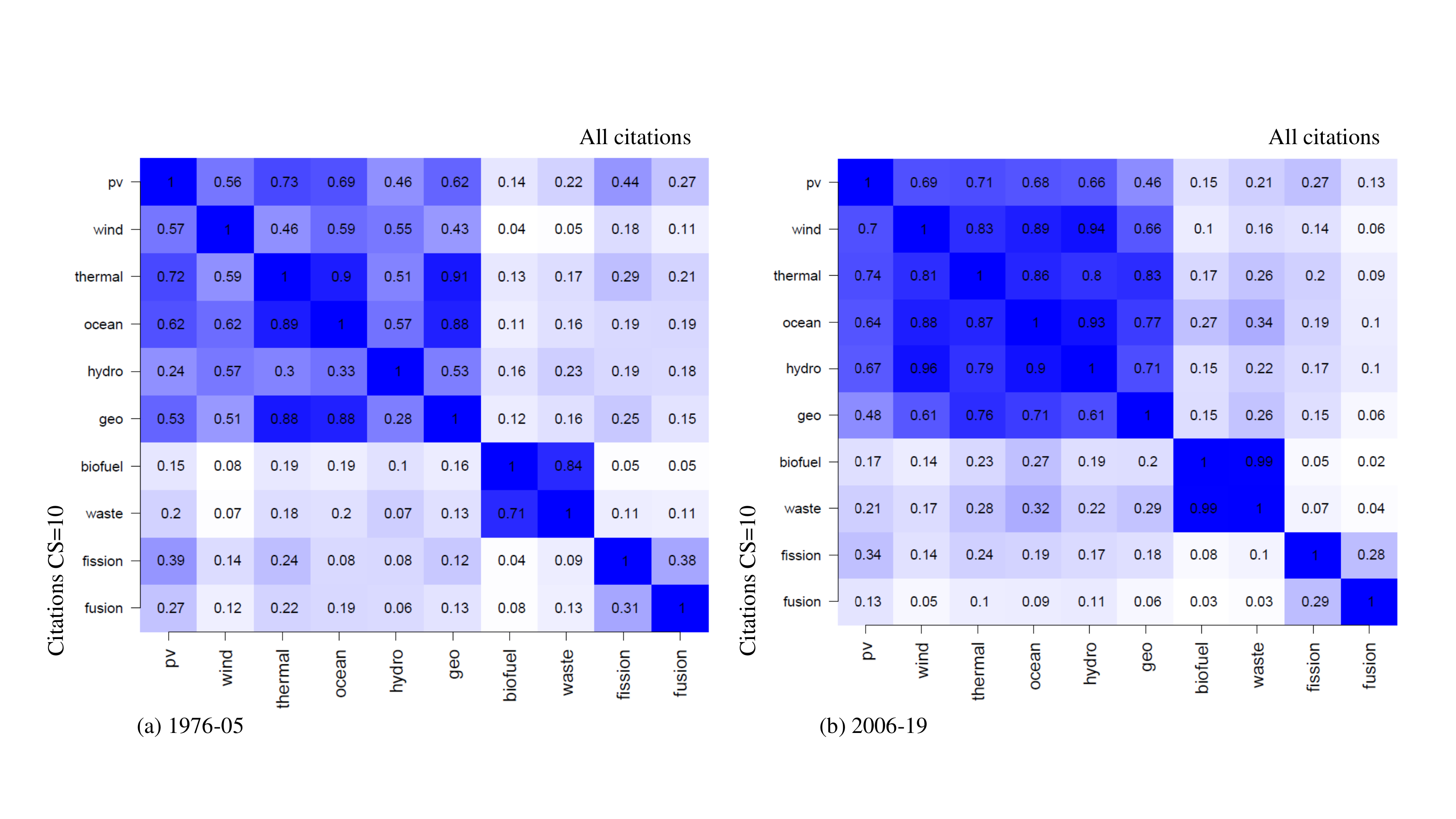}
\caption{
Robustness check for confidence scores.
Cosine similarities of scientific reliance based on citations to Web of Science fields for the time horizons (a) 1976-2005 and (b) 2006-2009.
The values in the upper triangle of the matrix are the cosine similarities obtained from using all available citations (as reported in the main text). In the lower triangle of the matrix cosine similarities are shown if only scientific citations with the highest confidence score $CS=10$ are considered. 
}
\label{fig:SI_robust_CS10_cosine}
\end{figure}

Patent-paper citation links in MAG are obtained by automated procedures.
The developers of RoS provide confidence scores ($CS\in\{1,10\}$) to indicate the reliability of the matched links where $CS=10$ indicates the highest confidence score.
Results shown in the main text are based on all citations which have confidence scores $>3$.

To check that results are not driven by citations of low confidence scores, parts of the analysis are redone with links of only highest confidence, $CS=10$.
When redoing Figs. \ref{fig:green_timeseries_count} and \ref{fig:pcs_share} for the highest confidence score citations only, results are qualitatively very similar to the figures presented in the main text, except that the level of ``scientificness'' overall decreases due to the omission of citations. We thus decided against plotting them again here.

Fig. \ref{fig:SI_robust_CS10_cosine} compares the values obtained from the network analysis in the main text with the values obtained from using only the highest-confidence citations. 
The upper triangles of the matrices show the cosine similarities which are also reported in the main text. Cosine similarities obtained from using only $CS=10$ citations are shown in the lower triangle. For most of the cases values are very similar and as the color-codes indicate, the qualitative patterns are essentially the same.

\subsection{Co-classification}
\label{sec:SI_robust_coclass}

As mentioned in the main text, multiple CPC technology codes can be assigned to a single patent. The frequency of technological codes appearing on the same patent is often used as an indicator for similarity of technologies (co-classification).
In this paper we analyze whether LCETs rely on similar knowledge sources, i.e. if they make similar citations to papers and patents. 
The cosine similarity networks considered here and co-classification networks are similar by construction: If two LCET categories are assigned to the same patent, citations made by the two LCETs would be the same when looking only at this single patent. 

\begin{figure}
\centering
\includegraphics[trim = {.75cm 1.75cm 0cm 2.8cm}, clip, width=1\textwidth]{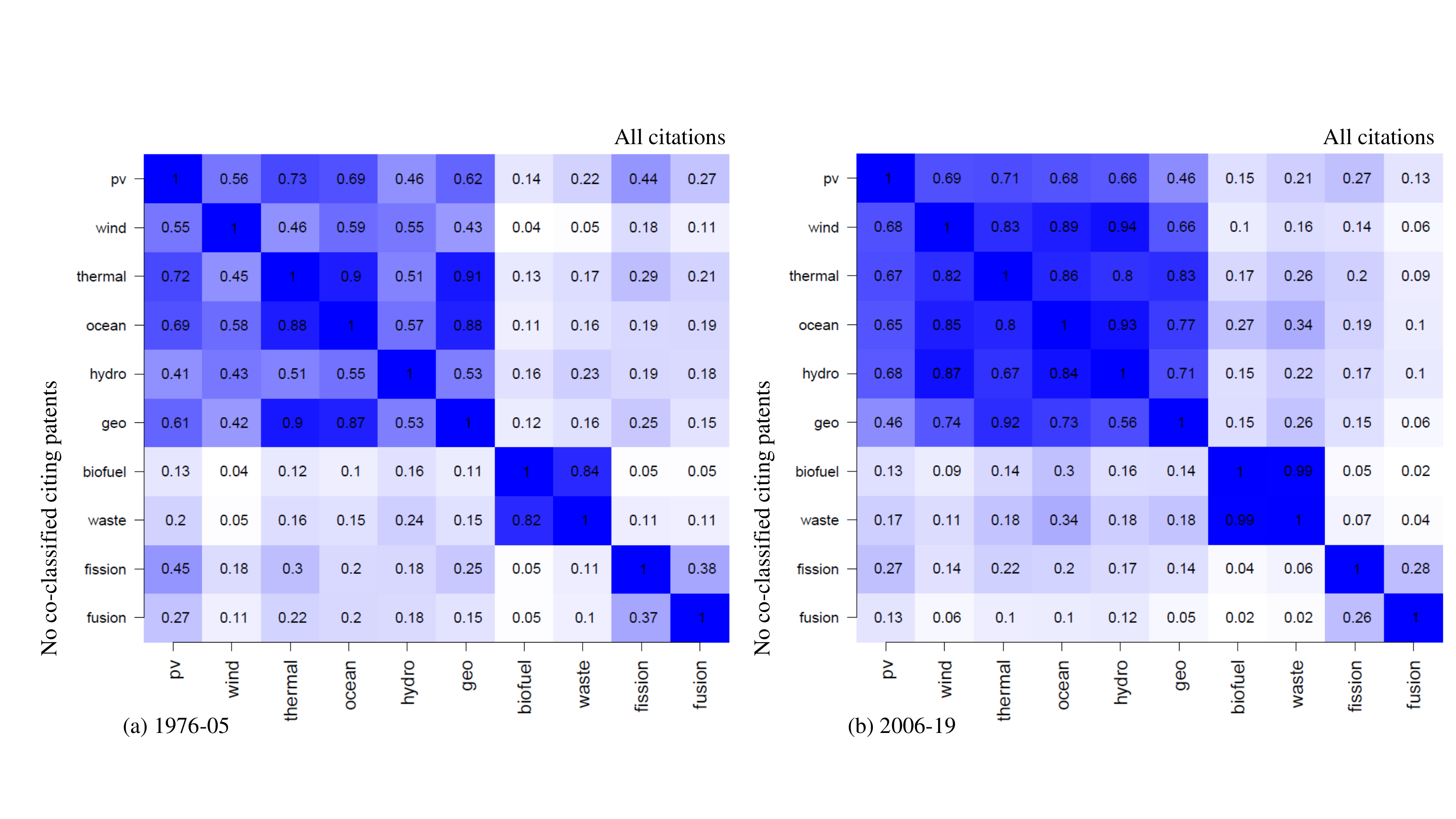}
\includegraphics[trim = {.75cm 1.75cm 0cm 2.8cm}, clip, width=1\textwidth]{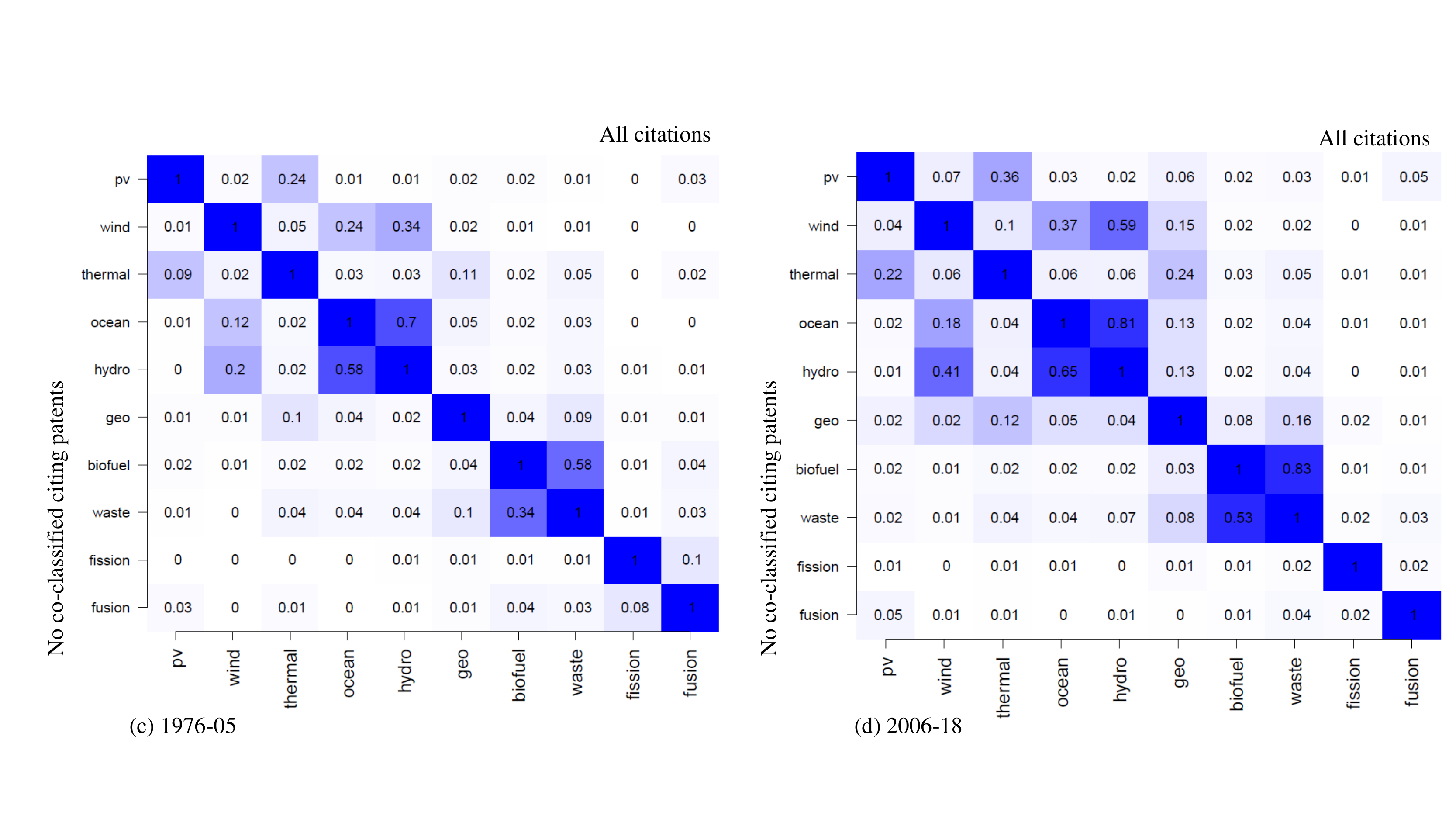}
\caption{
Robustness check for co-classification.
Cosine similarities of scientific reliance based on citations to Web of Science fields for the time horizons (a) 1976-2005 and (b) 2006-2009. (c) and (d) show cosine similarities based on citations to other patent CPC 4-digit categories.
The values in the upper triangle of the matrices are the cosine similarities obtained from using all available citations (as reported in the main text). The cosine similarity values shown in the lower triangle of the matrix are obtained after excluding all co-classified LCET patents. 
}
\label{fig:SI_robust_coclass_cosine_wos}
\end{figure}

To check that our knowledge base similarities are not driven by co-classification, we redo the technology network analysis after removing all patents which have multiple LCETs assigned.
Fig. \ref{fig:SI_robust_coclass_cosine_wos} shows the results of the robustness check. We again show cosine similarities reported in the main text on the upper triangle and the ones after excluding co-classified patents in the lower triangle.
Results are very robust against the removal of co-classified patents. This holds true for the technology-science networks Fig. \ref{fig:SI_robust_coclass_cosine_wos} (a)-(b) as well as for the technology-technology networks (c)-(d).

\FloatBarrier
\end{document}